\documentclass[final]{IEEEtran}
\usepackage{epsfig}
\usepackage{indentfirst,algorithmic}
\usepackage[english]{babel}
\usepackage{url}
\usepackage{graphicx,epstopdf}
\usepackage[ddmmyyyy,hhmmss]{datetime}
\usepackage{commath}
\usepackage{indentfirst,algorithmic}
\usepackage{amsfonts}
\usepackage{comment}
\usepackage{color}
\usepackage{mathtools}
\usepackage{amsmath}
\usepackage[colorinlistoftodos]{todonotes}
\usepackage{subcaption}
\usepackage{algorithm}
\usepackage{algorithmic}
\usepackage{indentfirst,algorithmic}
\usepackage{flushend}
\usepackage{float}
\usepackage{amsmath}
\usepackage{amssymb}
\usepackage{balance}
\usepackage{bm}
\usepackage{amssymb}
\usepackage[noadjust]{cite}

\begin{document}
\title{Estimating the Relative Speed of RF Jammers in VANETs} 
\author{\IEEEauthorblockN{Dimitrios Kosmanos, Antonios Argyriou and Leandros Maglaras }
\thanks{
Corresponding author: D. Kosmanos (email: dikosman@uth.gr).} 

\thanks{D. Kosmanos and Antonios Argyriou are in the Department of Electrical \& Computer Engineering,  University of Thessaly, Volos, Greece}
\thanks{L. Maglaras is in the Department of Computing Technology, De Montfort University, Leicester, UK.}
}
 \vspace{-5mm}
\maketitle

\noindent

\markboth{Latest version \today~at~ \currenttime}{IEEE TVT, Latest version \today~at~ \currenttime}

\begin{abstract}
Vehicular Ad-Hoc Networks (VANETs) aim at enhancing road safety and providing a comfortable driving environment by delivering early warning and infotainment messages to the drivers. Jamming attacks, however, pose a significant threat to their performance. In this paper, we propose a novel Relative Speed Estimation Algorithm (RSEA) of a moving interfering vehicle that approaches a Transmitter ($Tx$) - Receiver ($Rx$) pair, that interferes with their Radio Frequency (RF) communication by conducting a Denial of Service (DoS) attack. Our scheme is completely sensorless and passive and uses a pilot-based received signal without hardware or computational cost in order to, firstly, estimate the combined channel between the transmitter - receiver and jammer - receiver and secondly, to estimate the jamming signal and the relative speed between the jammer - receiver using the RF Doppler shift. Moreover, the relative speed metric exploits the Angle of Projection (AOP) of the speed vector of the jammer in the axis of its motion in order to form a two-dimensional representation of the geographical area. This approach can effectively be applied both for a jamming signal completely unknown to the receiver and for a jamming signal partly known to the receiver. Our speed estimator method is proven to have quite accurate performance, with a Mean Absolute Error (MAE) value of approximately $10\%$ compared to the optimal zero MAE value under different jamming attack scenarios. 
\end{abstract}

\section{Introduction}
Autonomous vehicles, capable of navigating in unpredictable real-world environments with little human feedback are a reality today \cite{platoons}. Autonomous vehicle control imposes very strict security requirements on the wireless communication channels that are used by a fleet of vehicles \cite{guardian},\cite{new-york-times}. This is necessary in order to ensure reliable connectivity \cite{vehicle-platooning}. Moreover, the Intelligent Vehicle Grid technology, introduced in \cite{ivg}, allows the car to become a formidable sensor platform, absorbing information from the environment, other cars, or the driver, and feed it to other vehicles and infrastructure so as to assist in safe navigation, pollution control and traffic management. The vehicle grid essentially becomes an Internet of Things (IoT) for vehicles, namely the Internet of Vehicles (IoV), that is capable of making its own decisions when driving customers to their destinations \cite{iov2017}. 

Wi-Fi has become essential for the operation of a modern vehicle \cite{wifi-car}. Wireless communications, however, being vulnerable to a wide range of attacks \cite{quyoom2015novel}. A RF jamming attack consists of radio signals maliciously emitted to disrupt legitimate communications. Such jamming is already known to be a big threat for any type of wireless network. With the rise in safety-critical vehicular wireless applications, this is likely to become a constraining issue for their deployment in the future. 
A subcategory of the jamming attacks is the Denial of Service (DoS) Attack, which is targeting to the availability of network services.
Of special interest are the mobile jammers, which impose an added strain on vehicular networks (VANETs). Thus, the accurate prediction of the behavior of the jammer such as its speed, becomes critical for providing a swift reaction to an attack. In this work, we propose a novel metric that captures the relative speed between the jammer ($Jx$) and the receiver ($Rx$). We also propose the Relative Speed Estimation Algorithm (RSEA) that is a completely sensorless and passive estimation method that uses pilot-based received signals at the receiver in order to, firstly, estimate the channel between the transmitter-receiver and jammer-receiver, secondly the jamming signal and thirdly, to estimate the relative speed between the jammer-receiver
using the RF Doppler shift property. This is the first work in the literature, according to our knowledge, that proposes an algorithm for speed estimation of malicious RF jammers.

\textbf{Problem Statement:} In addition to RF jamming, wireless communication between a transmitter ($Tx$) and a receiver ($Rx$) can be impaired by unintentional interference and multiple access control (MAC) protocol collisions. Jammers can exhibit arbitrary behavior in order to disrupt and thwart communication with a form of Denial of Service (DoS) attack \cite{YOUSAF2017124}. In the general case,  RF jamming reduces the receiver signal to interference and noise ratio (SINR), a problem that can be addressed with classic communication algorithms. However, in several applications it is critical to detect accurately the presence of a jammer, i.e. the precise reason behind the reduction in SINR, the packet-delivery-ratio (PDR), and more importantly, the nature of the attack. Consequently, it is difficult to determine whether the reason for the SINR reduction is an intentional jamming attack or unintentional interference. 
The challenge in detecting an RF jamming attack is that the information that is available for a jammer is typically minimal and derives from the useful signal possibly mixed with other types of arbitrary interference in the area.  Estimating the relative speed between a legitimate vehicle and a jammer, we can conclude if a high interference scenario has been provoked intentionally with the form of a DoS attack by an attacker that approaches the victim or has been provoked unintentionally by an area with significant RF interference. Specifically, if the estimated relative speed metric is about 0, we can conclude that the jammer is moving with about the same speed with the receiver, having as result the jammer to approach the receiver at some time instant. On the other hand, if the estimated relative speed gets a bigger quantity, we can conclude that the jammer is moving at quite different speeds than this of receiver. Of particular interest is the higher level behavior of a jammer, like its motion/movement relative to the $Tx$ - $Rx$ pair. This information can be effectively utilized in order a moving DoS attacker to be classified using Machine Learning algorithms.
\begin{figure}
\centering
 \includegraphics*[keepaspectratio,width =1.0\linewidth]{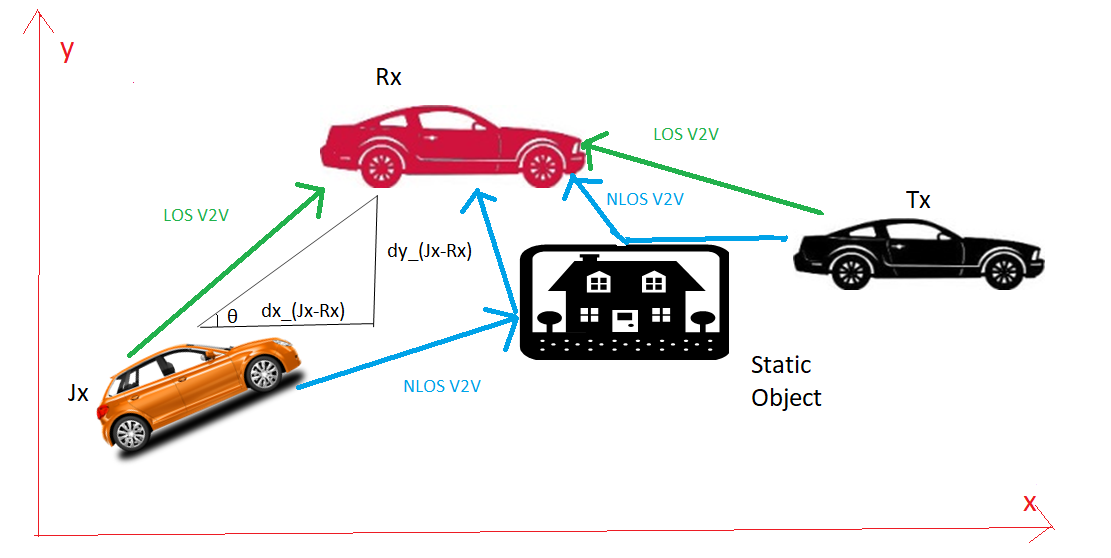}
\caption{ The Network Topology in which the orange vehicle is the jammer that approaches the $Tx - Rx$ pair from a AOP $(\theta)$ angle. This figure also includes the multipath fading effects by a static object}
\label{fig:doopler}
\end{figure}

\textbf{Our solution:} 
Using the jamming signal at the receiver we estimate the relative speed metric ($\Delta{u}$) that is based on the difference or sum between the velocities of the jammer and the receiver. This passively estimated metric also includes information regarding the Angle of Projection (AOP) of the jammed signal. Our scheme uses only the signal at the receiver under the presence of a jammer to characterize the behavior/motion of the jammer (if the $Jx$ is approaching or moving away from the $Rx$) using the RF Doppler shift. We also adopt a pilot-based method for the channel estimation between $Tx$ - $Rx$ since it is suitable for fast varying channels, such the VANETs, because the channel is directly estimated by training symbols or the pilot tone that are known a priori to the receiver. 

The contributions of the paper are three-fold:

\begin{itemize}
\item A completely sensorless and passive pilot-based scheme is proposed that is based on RF communication between $Tx - Rx$ being interfered by a jammer in the area. However, we do not apply the proposed RSEA for estimating only speed of $Tx$ \cite{velocity2009cell}. 
We try from this point-to-point communication to gather as much information as possible regarding jammer's behavior, such as all the combined multipath channels among $Tx, Rx, Jx$, jammer's relative speed value and the jamming signal; 

\item In addition, the proposed relative speed metric defines physical location features, because is combined with the AOP of the jammer;
 \item The effective usage of the estimated relative speed for a future jamming detection algorithm is outlined;
 \end{itemize}
 
It has to be highlighted that the proposed RSEA can also be applied with a completely unknown jamming signal, in the case where the $Tx$ sends more pilot symbols to the Rx than the sum of the number of different unknown to the receiver jamming symbols being sent by the jammer with the specific value of parameter $2N$, which is the double number of multipath rays in the area for the estimation of both two channels between $Tx-Rx$ and $Jx-Rx$. \par
The rest of this paper is organized as follows: Section \ref{related_work}
presents the related work, whilst Section \ref{system-model} analyzes the network topology, the system model analysis and the wireless channel model. Section \ref{location-aware} presents the location aware relative speed metric and Section \ref{proposed_algorithm} analytically describes the proposed RSEA. Section \ref{performance_evaluation} presents the experimental evaluation of the proposed RSEA and provides comparison between different scenarios. Finally, Section \ref{conclusion} concludes the paper and gives some directions for future work.

\section{Related Work}
\label{related_work}
\subsection{RF Jamming}
RF jamming has been extensively studied in the
context of classical 802.11 networks without accounting for
the particularities of car-to-car communications. Besides the
differences in PHY design of 802.11p compared to other
802.11 amendments, the propagation conditions of VANET
are fundamentally different due to the highly dispersive and
rapidly changing vehicular environment. A lot of kinds of jamming attacks has been studied in VANETs \cite{jamming-survey}. The two most important kinds of jamming attacks are the constant jamming and the reactive jamming. Constant jamming transmits random generated data on the channel without checking the state of the channel (Idle or not). However, the reactive jamming jams only when it senses activity on the channel otherwise it stays idle. In \cite{rf-jamming-punal} observe that constant, periodic, but also reactive RF jammer can hinder communication over large propagation areas, which would threaten road safety. Reactive jamming attacks reach a high jamming efficiency
and can even improve the energy-efficiency of the jammer in
several application scenarios \cite{reactive2},\cite{reactive3}. Also, they can easily and
efficiently be implemented on COTS hardware such as USRP
radios \cite{reactive1},\cite{reactive4},\cite{reactive5}. But, more importantly, reactive jamming attacks
are harder to detect due to the attack model, which allows
jamming signal to be hidden behind transmission activities
performed by legitimate users \cite{reactive4},\cite{reactive6},\cite{reactive7} . A different kind of attacks are the pilot-based attacks against
OFDM and OFDMA signals \cite{ofdm-pilot-jamming}. These attacks seek to manipulate
information used by the equalization algorithm to cause errors
to a significant number of symbols. However, we do not evaluate this type of attack because the point of interest of this paper is the DoS attacks that are targeting to availability and no to integrity. In order to be robust against pilot tone jamming attacks, OFDM and OFDMA systems must randomize their subcarrier
locations and values. For the mitigation of this type of RF jamming attack optimal power allocation with user scheduling are proposed under reactive jamming in the area \cite{power-allocation}, utilizing also the technique of uncoordinated frequency hopping (UFH) \cite{uhf-infocom}. UFH implies the communication between transmitter and receiver through a randomly
chosen frequency channel unknown for both agents. In \cite{uhf-friendly-jammers} in order the secrecy level of wireless networks
under UFH to be characterized, showing the harmful security effect of broadband
eavesdropper adversaries capable of overhearing in multiple
frequencies. To  be countered such eavesdroppers, we consider the
use of broadband friendly jammers that are available to cause
interference on eavesdroppers. The goal is to cause as much interference as possible to eavesdroppers that are located in unknown positions, while limiting the interference observed by the legitimate receiver. However, the information about the location and speed of frienly jammers are crucial for the above UHF schemes. 
\subsection{Localization}
 A lot of work has covered matters of localization, which is a fundamental challenge for any wireless network of nodes, in particular, when nodes are mobile. In \cite{velocity-estimation} the relative positions and velocities (PVs) are estimated up to a rotation and translation of an anchor-less network of mobile network, given two-way communication capability between all the nodes. A least squares based dynamic ranging algorithm is proposed, which employs a classical Taylor series based approximation to estimate pairwise distance derivatives efficiently without the usage of Doppler shifts. In \cite{speed-etc}, authors propose a dual-level travel speed calculation model, which is established under different levels of sample sizes. Wireless sensor networks (WSNs) are widely used to maintain the location information and rely on the tracking service only when their location changes. In the proposed approach in \cite{speed-doopler}, the problem of tracking cooperative mobile nodes in wireless sensor networks is addressed with the Doppler shifts of the transmitting signal in combination with a Kalman filter, by performing a constrained least-squares optimization when a maneuver is detected. In \cite{speed-estimation}, authors suggest a method for joint estimation of the speed of a vehicle and its distance to a road side unit (RSU) for narrow-band
orthogonal frequency-division multiplexing (OFDM) communication systems. Spatial filtering and a Maximum Likelihood (ML) algorithm is developed for distance estimation.
The vehicle speed is calculated using a kinematics model based on the estimated distance and Angle of Arrival (AOA) values.
\subsection{Speed Estimation}
Another class of papers proposes speed estimation systems that alerts drivers about driving conditions and helps them avoid joining traffic jams using multi-class classifiers. ReVISE in \cite{reVISE} proposes a multi-class SVM approach that uses features from the RF signal strength. Using a similar method, MUSIC \cite{MUSIC} is a subspace based AOA estimation algorithm that exploits the eigen-structure of the covariance matrix of the received signals on a multi-signal classifier.

Covariance-based speed estimation schemes have also been used for the estimation of the maximum Doppler spread, or equivalently, the vehicle velocity that, is useful for improving handoff algorithms \cite{speed-spectral}. The authors in \cite{mobile-speed} proposed an algorithm that employs a modified normalized auto-covariance of received signal power to estimate the speed of mobile nodes. 

However, the above covariance-based speed estimator utilizes correlation lags to improve performance which comes at the expense of computational complexity. In \cite{speed-mobile-phones} an algorithm that estimates the speed of a mobile phone by matching time-series signal strength data to a known signal strength trace from the same road is introduced. The drawback of the correlation algorithm is the observation that the signal strength profiles along roads remain relatively stable over time. However, the results are more accurate than previous techniques that are based on handoffs or phone localization. In \cite{speed-mobile-phones-new} a method for the estimation of speed for mobile phone users using WiFi Signal-to-Noise Ratio (SNR) and time-domain features like mean, maximum, and auto-correlation is proposed. Whilst in \cite{tracking} two novel autocorrelation (ACF) based velocity estimators are used, without requiring knowledge of the SNR of the link. 

In all the prior work, speed estimators that have been proposed include training procedures in order to estimate traffic congestion or other transportation performance metrics using sensor measurements.  However, speed estimation problem from wireless RF communication due to security issues has been not widely investigated. Only, in \cite{angle-of-arrival} the authors try to estimate the AOA of the specular line of sight (LOS) component of signal received from a given single antenna transmitter using a predefined training sequence.  The results show the optimality of the training based Maximum Likelihood (ML) AOA estimator in the case of a randomly generated jamming signal. However, the drawback of this ML-AOA estimator is that superior performance is subject to the availability of a perfect CSI. On the contrary, authors in \cite{doopler-shift} introduced a new algorithm to estimate the mobile terminal speed at base station in cellular networks. This helps BTS in estimating the channel Doppler shift, using  measured received signals at the Base Station (BS). The Doppler shift estimation algorithm is improved by utilizing a speed estimation window that slides over bursts with overlaps and by introducing two different low and high thresholds for power level comparisons. The performance of the proposed algorithm is modeled in a Terrestrial Trunked Radio (TETRA) network and the simulation has shown acceptable results for a wide range of velocities and jammers. However, there has been no prior work that combines a feature of the physical location, such as the AOP of the $Jx$ with its speed in order to estimate relative speeds of two moving vehicles (jammer - receiver)  during a jamming attack, using the channel Doppler shift value.

The great majority of the  works have used speed estimation in order to improve handover algorithms between transmitter and receiver under a typical micro-cellular system with non-isotropic scattering \cite{velocity2009cell} and for calculating the optimal tuning of parameters for systems that adapt to changing channel conditions.
Our proposed technique is the first in the literature, to the best of our knowledge, that uses the unicast communication between $Tx$ and $Rx$ for the prediction of the jammer's speed and for future detection of a jamming attack. 

\section{System-Model \& Preliminaries}
\label{system-model}
\subsection{ Network Topology}
We consider unicast V2V communication between transmitter and the receiver and a point-to-point V2V communication between a single jammer and the receiver. This simple scenario in a rural area is used for the initial verification of our system without high interference of other vehicles. In this area, a static obstacle already exists that impacts the communication between $Tx - Rx$ and $Jx - Rx$.

The jammer transmits wireless packets/signals that may form a reactive jamming signal. We assume that the $Tx - Rx$ pair of vehicles in our model moves with a constant speed for a period of time. 
This approach allows the modeling of platoons of vehicles that are formed by maintaining a constant distance with each other \cite{platoons}. We assume that the jammer moves with a constantly increasing speed with the ultimate goal to approach the receiver and intervene in the effective communication zone of the $Tx - Rx$ pair. As it can be seen in the network topology of Fig.\ref{fig:doopler}, the distances between $Jx - Rx$ in the y axis $dy_{(Jx-Rx)}$ and in the x axis $dx_{(Jx-Rx)}$ together with the actual distance between $Jx - Rx$ $d_{(Jx-Rx)}$, which is the hypotenuse of the rectangular triangle that is formed. The motion of the vehicles in Fig.\ref{fig:doopler} is characterized by the speed vectors $(\vec{u}_\text{Rx},\vec{u}_\text{Jx},\vec{u}_\text{Tx})$. Only $\vec{u}_\text{Rx},\vec{u}_\text{Tx}$ have the same direction, which is the direction of the x axis. The jammer approaches the $Tx - Rx$ with an AOP $(\theta)$. So, the speed vector of the jammer is projected in the axis of the motion of vector $\vec{u}_\text{Rx}$ with an AOP $(\theta)$  Fig.\ref{fig:projections_LOS}. In this figure we also notice that the AOP $(\theta)$ is not equal to zero. Moreover, the angle that is formed between the speed vector of the jammer $\vec{u}_\text{Jx}$, and the wireless signal that travels between the $Jx$ and the $Rx$, is called the Angle of Departure (AOD) and is denoted as $\phi$ in Fig.\ref{fig:projections}. In Fig.\ref{fig:projections_LOS} the Line of Sight (LOS) component between $Jx$ - $Rx$ has a AOD $(\phi)$ equal to zero, while the non Line of Sight (NLOS) component between $Jx$ - $Rx$ has a AOD $(\phi)$, which is different to zero in Fig.\ref{fig:projections_NLOS}.

\subsection{System Overview}
\label{system-overview}
In our system model, $K$ known pilot symbols that compose the symbol vector $\vec{x}_{pilot} = [x_{pilot}(1)...x_{pilot}(K)]^{T} = [1...1]^{T}$ are being sent over consecutive $K$ time instants from the transmitter to the receiver. At the same time, the jammer simultaneously transmits over consecutive $K$ time instants $K$ jamming symbols to the receiver that compose the symbol vector $\vec{s}=[s_1...s_K]^{T}$. So we consider the received vector at the receiver $\vec{y}=[y(1)y(2)...y(K)]^{T}$, which consists of the combined symbols that the $Rx$ receives from the transmitter and the jammer at $K$ consecutive time instants. Therefore, for every time instant $n\in (0,K]$ the receiver signal $y(n)$ is the summation of the pilot symbol sent by the transmitter $x_{pilot}(n)$ and the symbol sent by the jammer $s_n$.
Using pilots, the LOS channel and the $N-1$ NLOS channels between $Jx - Rx$ are estimated by the receiver. The receiver can also define the specific value of parameter $N$, which is the total numner of multipath rays. The wireless channel is assumed to be constant for the duration of the transmission of the $K$ pilot symbols from $Tx$ to $Rx$. 

\subsection{Attacker Model}
\label{attack}
We consider jammers that aim to block completely the communication over a link by emitting interference reactively when they detect packets over the air, thus causing a Denial of Service (DoS) attack. The jammers minimize their activity to only a few symbols per packet and use minimal, but sufficient power, to remain undetected. We assume that the jammer is able to sniff any symbol of the packet over the air in real-time and react with a jamming signal that flips selected symbols at the receiver with high probability (see \cite{detection-reactive}).\par

 The jammer is designed to start transmitting upon sensing energy above a certain threshold in order a reactive jamming attack to be succeed. We set the latter to $-86 dBm$ as it is empirically determined to be a good tradeoff between jammer sensitivity and false transmission detection rate, when an ongoing 802.11p transmission is assumed. So, the symbol vector $\vec{s}$ that reaches at the $Rx$ from the $Jx$ after $K$ time instants has the same length as the pilot symbol vector that reaches the $Rx$ from the $Tx$ after $K$ time instants, provided that the jammer transmits only when senses a transmission from the transmitter. Each one of the $K$ scalar values depends on the used power by the jammer. The jammer continuously transmits with
the same transmission power, with the purpose of overloading the
wireless medium representing, thus a DoS attack \cite{ddos-attack}.
This work assumes that when the jammer continuously transmits the same jamming symbol to the receiver forming a \textit{simplified jamming signal} of the form $\vec{s}= [f...f]^{T}$ with length $K$ and $f$ a random unknown value to the receiver. Furthermore, the proposed RSEA can operate with completely unknown jamming signals. This is possible when the $Tx$ sends more pilot symbols to the Rx than the sum of the different unknown jamming symbols being sent by the jammer.

Recall that the main goal of this paper is to show how we can estimate the speed of an non-cooperative malicatious attacker that can eventually be used as extra useful information for the design of a RF jamming detection schemes~\cite{karagiannis}.

\subsection{Channel Model}
\label{model: B}
\begin{figure}
\centering
\begin{subfigure}{.5\textwidth}
   \includegraphics[width=.8\linewidth]{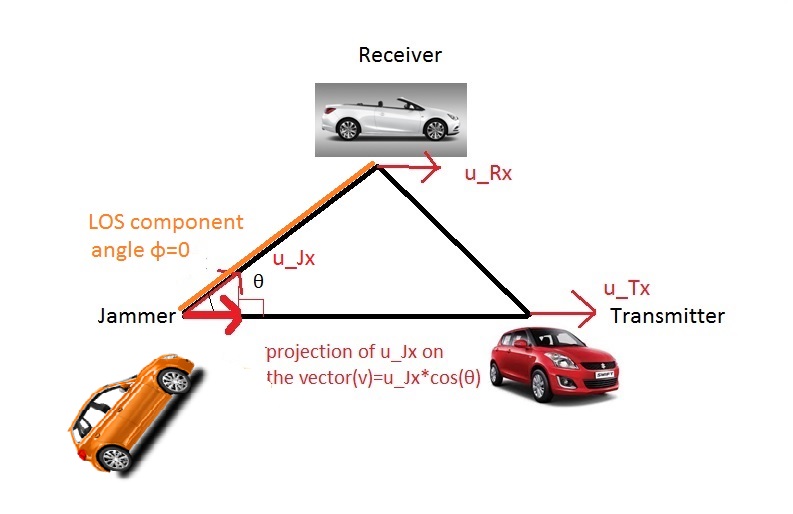}
  \caption{LOS ray of $Jx$-$Rx$ communication with $\phi=0$, $\theta\neq0$}
  \label{fig:projections_LOS}
\end{subfigure}
~
\begin{subfigure}{.5\textwidth}
  \centering
  \includegraphics[width=.98\linewidth]{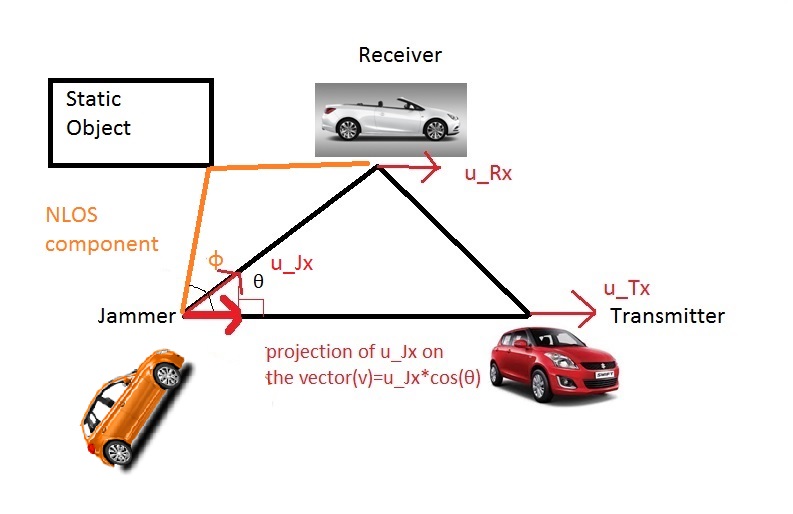}
  \caption{NLOS ray of $Jx$-$Rx$ communication with $\phi\neq0$, $\theta\neq0$}
 \label{fig:projections_NLOS}
\end{subfigure}
\caption{ Illustration of projections of velocities $u_\text{Jx}$ on the vector $\vec{v}$. Two-dimensional scheme}
\label{fig:projections}
\end{figure}

Multipath is the propagation phenomenon that results in radio signals reaching the receiving antenna by two or more paths.
The multipath scenario illustrated in Fig.\ref{fig:doopler} includes a static obstacle in order for the multipath effects to be considered in the communication between $Tx$ - $Rx$ and $Jx$ - $Rx$. So, it exists the LOS component of the wireless signal being sent by the $Jx$, $Tx$ and also the NLOS component. In the NLOS component the AOP $(\theta)$ is not equal to zero and the AOD $(\phi)$ between the speed vector of the jammer and the NLOS ray is also not equal to zero (see Fig.\ref{fig:projections_NLOS}). The phenomenons of reflection, diffraction and scattering due to the multipath give rise to additional radio propagation paths beyond the direct optical LOS path between the radio transmitter and receiver.\par
In our work, we adopt the Rician fading model, which is a channel model that includes path loss and also Rayleigh fading \cite{tse}. When a signal is transmitted the channel adds Rician fading. The
Rician fading model is particularly appropriate when there is a  direct propagating LOS component in addition to the faded component arising from multipath propagation.  

The Rician channel at time instant $t$ is defined with the help of multiple NLOS paths, which is similar to the Rayleigh fading channel but with the addition of a strong dominant LOS component. Parameter $q$ defines the channel between $Tx - Rx$ with $q=1$ and the channel between $Jx - Rx$ with $q=2$. We define a complex Gaussian random variable $\zeta_{G}$ that is uniform over the range $[0, 2\pi]$ and is fully specified by the variance $\sigma_{q}^{2}$. The Rician fading channel can be defined with the help of this random variable as:

\begin{equation}
\begin{aligned}
\label{eq:model_B}
h_{q}[t] &= \sqrt{\frac{k}{k+1}}\sigma_{q} e^{j(2\pi/\lambda ) (f_c+ f_{d,max} \cos{\phi_{q}}) \tau_{q} } \delta{(t-\tau_{q})} \\
& +\sqrt{\frac{k}{k+1}}\zeta_{G}
\end{aligned}
\end{equation}

In the above equation, $f_c$ is the carrier frequency, $f_{d,max}$ is the maximum Doppler shift, $\phi_{q}$ the incidence AOD between the vector of speed $\vec{u}_{Jx}$ with the vector of the jamming signal, $\tau_{q}= d/c$ is the excess delay time 
for the LOS ray that travels between the two communicating nodes in channel $h_q$, $d$ corresponds to the distance between the two communicating nodes and $t$ is the current time instant. The first term corresponds to the specular LOS path arrival and the second, to the aggregate of the large number of reflected and the scattered paths. Parameter $k$ is the ratio of the energy in the specular path to the energy in the scattered paths; the larger
$k$ is, the more deterministic the channel is \cite{multipath-fading}. Finally, $(\gamma_{q}=\sqrt{\frac{k}{k+1}}\sigma_{q})$ in \eqref{eq:model_B} is the complex amplitude associated with the LOS path, which is known at the receiver. Rician channel model is often a better model of representing fading compared to the Rayleigh model.

 The channel response $\vec{y}$ after $K$ consecutive symbols sent by the jammer and the transmitter is:
\begin{equation}
\begin{aligned}
\label{eqn:multipath}
\vec{y}&=\sum_{l=0}^{N-1}{(h_1[l]\vec{x}_{pilot}[N-l]+h_2[l] \vec{s}[N-l])} +\vec{w}
\end{aligned}
\end{equation}
In above equation, the $\vec{y}$ is a $K\times 1$ column vector. Moreover, $\vec{x}_{pilot}$ is the symbol vector that the $Rx$ receives from the $Tx$ after $K$ consecutive time instants and $\vec{s}$ is the symbol vector that the $Rx$ receives from the $Jx$ after $K$ consecutive time instants again. The symbol vectors $(\vec{x}_{pilot}[N-l]$,$\vec{s}[N-l])$ have the same values, as defined above, for the $l$ different paths of the respective channels, where $~\forall l\in (0,N-1]$. The $\vec{w}$ represents the additive white Gaussian noise (AWGN) with zero mean.
We assume, also, that the jammer and the transmitter send at very close time instants their symbols at the receiver, so that $h_1, h_2$ channels can remain stable for sending K symbols. Moreover, $N$ is the overall multipath rays  in the area. For the estimation of this parameter, we use the GEMV simulator \cite{GEMV-BOban}. For describing the modeled area GEMV uses the outlines of vehicles, buildings and foliage. Based on the outlines of the objects, it forms  R-trees. R-tree is a tree data structure in which objects in the field are bound by rectangles and are hierarchically structured based on their location in space. Hence, GEMV employs a simple geometry-based small-scale signal variation model and calculates the additional stochastic signal variation and the number of diffracted and reflected rays based on the information about the surrounding objects. We must note that the wireless RF communication of the $Tx$ - $Rx$ pair and the $Jx$ - $Rx$ pair is taking place in a specific frequency band, according to the existing standard for automotive systems\cite{wifi-car}. 

\subsection{ Transmission in the MAC/PHY Layer}
\label{mac-pilot}
We assume single carrier communication at the PHY. The 802.11p MAC also provides prioritized Enhanced Distributed Channel Access (EDCA), and can support applications by providing different levels of Quality of Service (QoS).
In our model, only the 802.11p MAC EDCA AC[0] channel with higher priority is used for the pilots. The pilot beacons from the $Tx$ to the $Rx$ are transmitted with high probability of successful delivery, increasing the accuracy of the proposed RSEA at the same time. Any type of collisions at the wireless channel
resulting from competing traffic is addressed by the MAC EDCA backoff mechanism for distances smaller than the Carrier Sensing (CS) range of $1000m$. So we assume that our speed estimation algorithm has a correct reaction and for high interference situations from other vehicles.

\section{Location Aware Relative Speed Metric}

One of the main novel ideas of this work, is that we take into account the physical location of the $Jx,Rx$ nodes and the direction of their motion when calculating the relative speed metric. In the general case, the $Rx$ does not move in the same direction as the $Jx$ (see Fig.\ref{fig:doopler}). For this case, $\Delta{u}$ includes the AOP (angle $\theta$) of the $Jx$ between $Jx$ and $Rx$. The geometry-aware metric takes into account the distance $dy_{(Jx-Rx)}$ and the distance $dx_{(Jx-Rx)}$. So a rectangular triangle is formed by the sides $dx_{(Jx-Rx)},d_{(Rx-Jx)},dy_{(Jx-Rx)}$. As it can be seen from Fig.\ref{fig:doopler}, the distance $d_{(Jx-Rx)}$ is the hypotenuse of the rectangular triangle, which means that $\cos{(\theta)}=dx_{(Jx-Rx)}/d_{(Jx-Rx)} $. So, the speed of the $Jx$ (Source) with respect to the $Rx$ speed, while the $Jx$ and the $Rx$ are moving in the same direction, is the relative speed between the two vehicles moving towards each other and is equal to the sum of their individual 
speed vectors $\Delta{\vec{u}_{line}}=\vec{u}_\text{Jx}+\vec{u}_\text{Rx}$. Moreover, $\vec{v}= \frac{\vec{u}_\text{Tx}}{||u_\text{Tx}||}$ is the unit length vector pointing from the $Jx$ to the $Tx$. 
The relative speed of the $Jx$ and the $Rx$ can be defined as the following dot product:

\begin{equation}
\label{metric}
\Delta{u}= \vec{v} \Delta{\vec{u}_{line}}
\end{equation}

To represent all the speed vectors of Fig.\ref{fig:doopler} in two dimensions $(x,y)$, we project the vector $\vec{u}_\text{Jx}$ on the unit length vector $\vec{v}$. The direction of $\vec{v}$ is the x axis (see Fig.\ref{fig:projections}). The projected vector is $\vec{u}_\text{Jx} \cos{(\theta)}$. On the other hand, $\vec{u}_\text{Rx}$ is already a vector in the direction of the x axis (see Fig.\ref{fig:projections}), which has the same direction as the projection of $\vec{u}_\text{Jx}$. This allows the calculation of the relative speed between $Jx - Rx$ using the two vectors $(\vec{u}_\text{Jx} \cos{(\theta)},\vec{u}_\text{Rx})$ that have the same direction with the vector $\vec{v}$.
\label{location-aware}

In \eqref{metric}, if we use the projection vector $\vec{u}_\text{Jx} \cos{(\theta)}$ and the $\vec{u}_\text{Rx}$ vector, we get the final version of our metric, which is:
\begin{equation}
\label{metric_new}
\Delta{u}=|\vec{u}_\text{Jx} (dx_{(Jx-Rx)}/d_{(Jx-Rx)}) + \vec{u}_\text{Rx} |=|\vec{u}_\text{Jx} \cos{(\theta)} + \vec{u}_\text{Rx} |
\end{equation}

This is the $\Delta{u}$ metric in the direction of $\vec{v}$. The addition is justified by the fact that the vectors $\vec{u}_\text{Jx} \cos{(\theta)}, \vec{u}_\text{Rx}$ have the same direction. In the above equation, $\vec{u}_\text{Jx}, \vec{u}_\text{Rx}$ are the speed vectors of the $Jx$ and the $Rx$, respectively. According to our model, if the $Jx$ approaches the receiver, $\cos{(\theta)}$ increases. As the $\vec{u}_\text{Rx}$ remains constant and the $\vec{u}_\text{Jx}$ is constantly increasing, \eqref{metric_new} is an increasing function and its maximum value indicates a nearby jamming attack.

As $\Delta{u}$ increases, the jammer approaches the receiver and when $\Delta{u}$ decreases, the jammer is moving away from the $Tx$ and the $Rx$. If the $Jx$ and the $Rx$ are located on the same road, an actual straight line and the vectors $\vec{u}_\text{Jx},\vec{u}_\text{Rx}$, have the same direction, then our metric is the sum between $Jx - Rx$ speed vectors ($\vec{u}_\text{Jx}+\vec{u}_\text{Rx})$. Otherwise, if the vectors $\vec{u}_\text{Jx},\vec{u}_\text{Rx}$ have opposite directions, our metric is estimated by  the difference ($\vec{u}_\text{Jx}-\vec{u}_\text{Rx})$. 

Taking into account the direction of the $Jx$ relative to the direction of the $Rx$ the general form of the above metric is:
\begin{equation}
\label{metric_new_general}
\Delta{u}=|\vec{u}_\text{Jx} \cos{(\theta)} \pm \vec{u}_\text{Rx} |
\end{equation}
It is crucial to point out that the above metric is the actual value of the relative speed that will be used in the subsequent sections to model the Doppler shift between the jammer and the receiver.

\section{Proposed Estimation Scheme}
\label{proposed_algorithm}

\subsection{Estimation of the Combined Pilot/Jamming Signal}
\label{h1_estimation}
The channel between two nodes with jamming is captured in \eqref{eqn:multipath}. For the proposed RSEA a pilot-based method for channel estimation is used. So, 
the signals that $Rx$ receives from the $Tx$ and the jammer interfere additively. In \eqref{eqn:multipath}, if we differentiate the one LOS component from the other $N-1$ NLOS components, we have:

\begin{equation}
\label{overall_eq}
\vec{r}_\text{LOS}= h_{1}^{LOS}\vec{x}_\text{pilot}[N] + h_{2}^{LOS}\vec{s}[N]
\end{equation}
Where, the channel values $h_{1}^{LOS}=h_{1}[0], h_{2}^{LOS}[2]=h_{2}[0]$ and the symbol vectors $\vec{x}_\text{pilot}[N], \vec{s}[N]$ represent the unique LOS component of the total $N$ multipath values.
If the NLOS multipath component is added:

\begin{equation}
\label{overall_eq1}
\vec{y}= \vec{r}_\text{LOS}+\sum_{l=1}^{N-1}{(h_1[l]\vec{x}_\text{pilot}[N-l]+h_2[l] \vec{s}[N-l])}+\vec{w}
\end{equation}

In \eqref{overall_eq1}, the received vector $\vec{y}$ is the convolution between $h_1$ and the pilot symbol vector $\vec{x}_{pilot}$ and the convolution between $h_2$ and the jamming symbol vector $\vec{s}$. Moreover, the $K\times 1$ column received vector $\vec{y}$ for the $K$ received values for every time instant during which the receiver collects every pilot that is sent from the transmitter is: 

\small
\begin{equation}
\begin{aligned}
\label{overall_eq2}
\vec{y}= \left[ \begin{array}{cc} r_\text{LOS}[1]+\sum_{l=1}^{N-1}{(h_1[l]+h_{2}[l]s_{1}[N-l])} \\...\\ r_\text{LOS}[K]+\sum_{l=1}^{N-1}{(h_1[l]+h_{2}[l]s_{K}[N-l])}
\end{array} \right] 
+ \left[\begin{array}{cc} w_1 \\...\\w_K \end{array} \right]
\end{aligned}
\end{equation}
\normalsize
To estimate the channel between $Tx$ - $Rx$ $(h_1)$, the channel between $Jx$ - $Rx$ $(h_2)$ and the jamming symbol vector $\vec{s}$, the best we can do is to estimate the combined vector parameter:

\begin{equation}
\label{eq:estimated}
\vec{z}= \left[ \begin{array}{cc} \sum_{l=0}^{N-1}{(h_1[l] x_\text{pilot}[N-l]+h_{2}[l]s_{1}[N-l])} \\...\\ \sum_{l=0}^{N-1}{(h_1[l] x_\text{pilot}[N-l]+h_{2}[l]s_{K}[N-l])}\end{array} \right]\\ \\
\end{equation}

 Vector $\vec{z}$ has the above form for the short time that is required by the receiver to collect all the $K$ symbols of the pilot vector. Recall that for a short time duration, the wireless channel is assumed constant. So for all the $K$ values of vector $\vec{z}$ in \eqref{eq:estimated}, the parameters $h_1[l],h_{2}[l],x_\text{pilot}[N-l]$ remain constant and only the jamming symbols may change depending on the form of the jamming symbol vector sent by the jammer. We use a MMSE estimator \cite{mmse}, which finds a better estimate from least squares (LS), in order the $K$ values of $\vec{z}$ to be estimated:
 \begin{equation}
\label{mmse}
\hat{\vec{z}}= (\vec{x}_{pilot}^{H} C_w^{-1} \vec{x}_{pilot})^{-1} \vec{x}_{pilot}^{H} C_w^{-1} \vec{y}
\end{equation}

$C_w$ is the covariance matrix of the noise vector $\vec{w}$. Vector $\vec{z}$ in \eqref{mmse} has $K$ components each having $N$ unknown multipath channel components. So, both the $h_1,h_2$ channels can be estimated and also the K values of the jamming signal $\vec{s}$ can be estimated too.

If the \textit{simplified jamming signal} is used\footnote{$\vec{s}=[f...f]^{T}$}, in which the jammer continuously sends the same jamming symbol, which is unknown to the $Rx$, we have $2N$ unknown values for the two channels $h_1,h_2$ with $K$ equations in \eqref{eq:estimated} and one unknown value for the jamming symbol $f$. So if the condition $K>2N+1$ is valid, we can see that each one of the channel values $h_{1},h_{2}$ out of $N$ multipath values can be estimated with the elimination method for the solution of the linear system with $K$ equations and $2N$ unknown values in \eqref{eq:estimated}. The values of the wireless channels $h_{1},h_{2}$ remain constant for each value of vector $\vec{z}$.
Moreover, the above linear system can also be solved with a completely unknown to the $Rx$ jamming signal, provided that the length of the pilot symbol vector $\vec{x}_{pilot}$ being sent from the $Tx$ to the $Rx$ is larger than the sum of the number of the unknown jamming symbols with the value of parameter $2N$, which is the double number of overall multipath rays in the area for the estimation of both $h_1, h_2$ channels.
We only utilize the LOS component of the vector $\vec{z}$ for the estimation of the relative speed metric using Doppler shift. So, the useful part from vector $\vec{z}$ that we need for the relative speed estimation through the Doppler shift is $\vec{r}_\text{LOS}=(\left[ \begin{array}{cc} h_{1}^{LOS}+ h_{2}^{LOS} s_{1} \\...\\ h_{1}^{LOS}+ h_{2}^{LOS} s_{K}\end{array} \right] )$. If we only want to estimate the $h_{1}^{LOS},h_{2}^{LOS}$ values of vector $\vec{r}_\text{LOS}$ without the multipath values, the above conditions for the solution of the linear system in \eqref{overall_eq2} can be simplified to $K>4$ for the \textit{simplified jamming signal} form. 
\subsection{Proposed Algorithm }
\begin{algorithm}
\caption{Relative Speed Estimation Algorithm (RSEA)}
\label{alg1}
\begin{algorithmic}[1] 
\STATE $N$  ~~\% It is specified by the $Tx$ for the specific area using the GEMV propagation model.
\FOR  {\textbf{Every time step $(t^{RSEA})$} A pilot signal with $K=2N+2$ symbols being sent from $Tx$ to $Rx$}  
  \STATE $N$  ~~\% It is re-specified by the $Tx$ for the specific area using the GEMV propagation model.
  \STATE $\hat{\vec{z}} \leftarrow$ MMSE($\vec{y},C_w^{-1}$)
\STATE $\vec{r}_\text{LOS} \leftarrow$ $(\left[ \begin{array}{cc} h_{1}^{LOS}+ h_{2}^{LOS} s_{1} \\...\\ h_{1}^{LOS}+ h_{2}^{LOS} s_{K}\end{array} \right] )$~~\%LOS components 
\IF{(($K > 2N+1$) ))~~\% and $\vec{s}$ has the \textit{simplified jamming signal} form}
\STATE  $\hat{\vec{r}}_\text{LOS}\leftarrow$ $(\left[ \begin{array}{cc} h_{1}^{LOS}+ h_{2}^{LOS}s \\...\\ h_{1}^{LOS}+ h_{2}^{LOS}s \end{array} \right] )$ ~~\% The $\vec{r}_\text{LOS}$ and $\vec{z}$ values can be estimated.
\STATE $\hat{r}_\text{LOS}[1]-h_{1}^{LOS}$ $=  (a_1 + b_1 j)$s
\ENDIF
\STATE $\hat{\Delta{u}}$ Estimation~~\% estimated relative speed value from (8)
\ENDFOR
\end{algorithmic}
\end{algorithm} 

The proposed RSEA is presented in Algorithm 1. First, the $Tx$ specifies the number of multipath rays $N$ in the area that the GEMV propagation model is used, as explained in subsection \eqref{system-model}. Then, the RSEA is used for every time step with the transmission of a pilot that consists of $K=2N+2$ symbols. In line 4 of the algorithm the combined channel between the $Tx$ and the $Rx$, with the intervention of the $Jx$, is estimated from the vector $\vec{y}$ using a MMSE estimator. Depending on the jamming signal, the inequality
that must be valid for the RSEA system to be resolvable for all the $N$ multipath values is different. In the final 10th line of the RSEA, the relative speed value is
estimated. A component $\hat{r}_\text{LOS}[1]$ of the estimated vector of the combined LOS channels $\hat{\vec{r}}_\text{LOS}$\footnote{Each one of the $K$ components of $\vec{r}_\text{LOS}$ has the same combined channel values} can be combined with the ray-optical baseband complex-number $(a_{1} +b_{1}j)s_{1}$, which is the jamming signal that the $Rx$ finally receives from the $Jx$. Specifically, the subtraction of the channel $h_{1}^{LOS}$ component from the $\hat{r}_\text{LOS}[1]$ value can be set equal to the ray-optical baseband complex-number $(a_{1} +b_{1}j)s_{1}$. The complex number $(a_{1} +b_{1}j)s_{1}$ characterizes the baseband form of the narrowband wireless channel. This narrowband wireless channel is a function of the relative speed $\Delta{u}$ between the jammer and receiver and the
Doppler shift between the two moving objects.

\subsection{Channel Model with Doppler Shift}
\label{doppler}
In this subsection, we describe in more detail the wireless LOS combined channel model $h_1^{LOS}+h_2^{LOS}$ between $Tx$ - $Rx$ and $Jx$ - $Rx$. The tracked LOS components also show fading characteristics, likely due to the ground reflection which cannot be resolved from the true LOS. For this reason, we choose
the same model for the LOS component as for the discrete components. So central to this paper, is the introduction of the proposed metric $\Delta{u}$ in the channel model of ~\eqref{eq:model_B}, taking into account the pathloss value at the receiver. This pathloss value only depends on the distance between the communicating nodes and usually gets small values for a narrowband wireless channel. Let us consider the channel model such as defined by the $Rx$ for a ray transmitted between two nodes as \cite{adaptive-model}:

\small
\begin{equation}
\begin{aligned}
\label{eqn:h_ij}
\sum_{q=1}^{2}{h_{q}^{LOS}(t,\tau_{q})}&= \\
& \sum_{q=1}^{2}\gamma_{q} {po_{q} e^{j(2\pi/\lambda ) (f_c+ f_{d,max} \cos{\phi_{q}}) \tau_{q} } \delta{(t-\tau_{q})}}
\end{aligned}
\end{equation}
\normalsize

In the above equation, $q$ defines the channel between $Tx - Rx$ with $q=1$ and the channel between $Jx - Rx$ with $q=2$, $\gamma_{qi}$ is the complex amplitude associated with the LOS path and $po_{q}$ represents the free space propagation loss \cite{path_loss}, $\lambda$ is the wavelength, $f_c$ the carrier frequency, $f_{d,max}$ is the maximum Doppler shift that depends on the $\Delta{u}$ metric such as in \eqref{metric}, $\phi_{q}$ is the incidence AOD between the vector of speed $\vec{u}_{Jx}$ and the vector of the jamming signal, $(\tau_{q}= d/c)$ is the excess delay time that the ray travels between the two nodes, and $t$ is the current time instant. We assume the LOS case for the communication between the jammer and the receiver, as can be seen in Fig.\ref{fig:projections_LOS}. The LOS ray between the $Jx$ and the $Rx$ has the same direction with the speed vector of the jammer. As a consequence the AOD is equal to zero ($\cos{\phi_{q}}=1$, in \eqref{eqn:h_ij}). The observed frequency at the receiver is $f'=f_c(1+\frac{\Delta{u}}{c} \cos{\phi_{q}}) $, which depends on the relative speed $\Delta{u}$ of the two vehicles (jammer, receiver) that we defined in the previous subsection. The baseband channel model for a ray transmitted between two nodes with the intervention of a jammer therefore becomes:

 \begin{equation}
 \begin{aligned}
 \label{h-channel}
\sum_{q=1}^{2}{h_{q}^{LOS}(t,\tau_{q})}=\sum_{q=1}^{2}\gamma_{q} {po_{q} e^{j(2\pi/\lambda ) f_c (1+ \frac{\Delta{u}}{c} \cos{\phi_{q}}) \tau_{q} }\delta{(t_i-\tau_{q})}}
\end{aligned}
\end{equation}

We can see that the Doppler shift $\Delta{f}$ Hz that is observed in the $Rx$ can be equal with \cite{stochastic1}:

\begin{equation}
\Delta{f}= \frac{\Delta{u} f_c \cos{\phi_{q}}}{c}
\end{equation}
And the maximum Doppler shift is:
\begin{equation}
f_{d,max}=  \frac{\Delta{u}}{c}
\end{equation}

Now, let $\tau_{q}$ be the time that is required for a signal to travel the distance $d$. Then, we can re-write $h_{2}^{LOS}$ from \eqref{h-channel} as:

\begin{equation}
\label{fc_h}
h_{2}^{LOS}(t,\tau_{2})=\gamma_{2} po_{2} e^{j(2 \pi/\lambda ) f_c (1 +  \frac{\Delta{u}}{c}) \frac{d}{c}}  \delta{(t-(\frac{d}{c}))}
\end{equation}

In the above equation, we use a $f_c=5,9Ghz$, which is the band dedicated to V2V communication. The channel $h_{2}^{LOS}(t,\tau_{2})$ is also the channel of a baseband signal in \eqref{fc_h} and if $(\frac{\Delta{u}}{c}>>1)$ has the form:

\begin{equation}
\label{c-sim}
h_{2}^{LOS}(t,\tau_{2})=\gamma_{2} po_{2} e^{j(2\pi/\lambda ) ( f_c \frac{\Delta{u}}{c}) \frac{d}{c}}  \delta{(t-(\frac{d}{c}))}
\end{equation}

To get our final signal model, we replace the path-loss parameter $po$ with equation:

\begin{equation}
\label{path-loss}
 po= G_{0,p} (\frac{d_{ref}}{d})^{n_p}
\end{equation}
Where, $G_{0,p}$ is the received power at a
reference distance $d_{ref}$, which is a standard value at about 100m, $n_p$ is the path-loss exponent, which is equal to 2 for the pure LOS links and $d_i$ is the distance that the transmitted ray travels between the two communicating nodes. So, $po$ only depends on the distance $d$ that the ray travels. We denote $\Delta{t}=t_{i}^{RSEA}-t_{i-1}^{RSEA}$ as the time interval between the current time instant and the preceding one, in which the RSEA is reapplied $(t_{i-1}^{RSEA})$. Furthermore, if $h_{2}^{LOS}(t,\tau_{2})$ represents the channel between the $Rx$ - $Jx$ pair, the distance between the two nodes after the time interval $\Delta{t}$ is $d=\Delta{u}\Delta{t}$, when the jammer approaches the receiver. Substituting (\ref{path-loss}) into \eqref{c-sim}, $h_{2}^{LOS}$ can be rewritten as: 
\small
\begin{equation}
\begin{aligned}
\label{po-sim}
 h_{2}^{LOS}(t,\tau_{2})&=\gamma_{2} G_{0,p} (\frac{d_{ref}}{ \Delta{u} \Delta{t}})^{2}  e^{j(2\pi/\lambda ) ( f_c \frac{\Delta{u}}{c}) \tau_{2} } \delta{(t-(\frac{d}{c}))   }
 \end{aligned}
\end{equation}
\normalsize

In the above equation, the only unknown parameter is $\Delta{u}$ at time $t$. Reorganizing \eqref{po-sim} we have:

\begin{equation}
\begin{aligned}
\label{euler}
h_{2}^{LOS}(t,\tau_{2})&= \gamma_{2} G_{0,p} (\frac{ d_{ref}^{2} }{ \Delta{u}^{2} \Delta{t}^{2}})  \delta{(t-(\frac{d}{c}))}( \cos{(\omega_{2})} \\
&+ j \sin{(\omega_{2})})
\end{aligned}
 \end{equation}
where, $\omega_{2}=(2\pi/\lambda ) (f_c \frac{\Delta{u}}{c} ) \tau_{2i} $.
In the above equation, the only unknown parameter is $\Delta{u}$.

For the LOS channel between $Tx - Rx$, we know that the receiver moves with the same speed as the transmitter, such as a platoon of vehicles with two members. The above means that the Doppler phenomenon is non-existent. Following \eqref{c-sim} for the formulation of the channel $h_{1}^{LOS}$ without the existence of Doppler phenomenon, we can see that this channel only depends on the path-loss component and the complex
amplitude associated with the LOS path. The path-loss component $po_{1}$ and the complex
amplitude variable $\gamma_{1}$ can be estimated by the receiver. So the $h_{1}^{LOS}$ can be represented by a complex number:

\begin{equation}
\label{h1-channel}
 h_{1}^{LOS}(t,\tau_{1})= \gamma_{1} po_{1}e^{0}= a_{Tx-Rx} + b_{Tx-Rx} j
\end{equation}

Reformulating the combined value of the LOS channels ($h_1^{LOS}$,$h_2^{LOS}$) in \eqref{h-channel} by combining equations \eqref{h1-channel}, \eqref{euler}, we have:

\small
\begin{equation}
\begin{aligned}
 \label{h-combined}
\sum_{q=1}^{2}{h_{q}^{LOS}(t,\tau_{qi})}&= \gamma_{1} po_{1}\\ 
&+\gamma_{2} G_{0,p} (\frac{ d_{ref}^{2} }{ \Delta{u}^{2} \Delta{t}^{2}})  \delta{(t-(\frac{d}{c}))}( \cos{(\omega_{2})} \\
&+ j \sin{(\omega_{2})})
\end{aligned}
\end{equation}
\normalsize

\subsection{Relative Speed Estimation}

At this point we have an estimate of the baseband channel $h_{2}^{LOS}$ between $Jx - Rx$, which can be represented with a complex number. The final baseband signal that reaches at the receiver after the intervention of the jammer can be represented as $(a_1 + b_1 j)s$. From Algorithm (1), we know that the jamming symbols of the symbol vector $\vec{s}$ is part of the vector $\vec{r}_\text{LOS}$. So, if from the estimated combined value $\hat{r}_\text{LOS}$ we subtract the channel $h_{1}^{LOS}$, which can be estimated by the receiver, the value $(\hat{r}_\text{LOS}-h_{1}^{LOS}=h_{2}^{LOS}$s) can be set equal to the baseband received signal at the receiver:

\begin{equation}
\label{base}
\hat{r}_\text{LOS}-h_{1}^{LOS}=(a_1 + b_1 j)s
\end{equation}
From the above equation, as well: 
\begin{equation}
h_{2}^{LOS}s = (a_1 + b_1 j)s
\end{equation}

Reusing the \eqref{euler} from the previous Section, the ray-optical baseband complex number $(a_1 + b_1 j)$ can be set equal with:

\small
\begin{equation}
\begin{aligned}
(a_1 + b_1 j)s&=\gamma_{2} G_{0,p} (\frac{ d_{ref}^{2} }{ \Delta{u} 
\Delta{t}^{2}} )\delta{(t-(\frac{d}{c}))}(\cos{(\omega_{2})}Re(s) \\
&+  j \sin{(\omega_{2})}Im(s) )
\end{aligned}
\end{equation}
\normalsize

where, $\omega_{2}=(2\pi/\lambda ) (f_c \frac{\Delta{u}}{c}) \tau_{2} )$.
The jamming signal $\vec{s}$ is estimated by the receiver from Algorithm (1). So the $Re(s),Im(s)$ are known values to the receiver. From the above equation, we can calculate the desired parameters $a_1,b_1$:

\small
 \begin{equation}
 \begin{aligned}
\label{eq1}
\small
a_1/ (\gamma_{2} G_{0,p} (\frac{ d_{ref}^{2}\delta{(t-(\frac{d}{c}))} }{ \Delta{u}^{2} \Delta{t}^{2}}) )= \cos{((2\pi/\lambda ) (f_c \frac{\Delta{u}}{c} ) \tau_{2})}
\end{aligned}
\end{equation}
\normalsize

\begin{equation}
\label{eq2}
b_1/ (\gamma_{2} G_{0,p} (\frac{ d_{ref}^{2}\delta{(t-(\frac{d}{c}))} }{ \Delta{u}^{2} \Delta{t}^{2}}) )= \sin{((2\pi/\lambda ) (f_c \frac{\Delta{u}}{c} ) ) \tau_{2})}
\end{equation}
 From \eqref{eq1},\eqref{eq2} and with the use of the Euler identity, we have:
 \begin{equation}
 \label{u_unknown}
\cos((2\pi/\lambda ) (f_c \frac{\Delta{u}}{c} ) ) \tau_{2})^{2}+ \sin ((2\pi/\lambda ) (f_c \frac{\Delta{u}}{c} ) ) \tau_{2})^{2}=1
 \end{equation}
In (\ref{u_unknown}) there is only one unknown variable $\Delta{u}$. So, we can calculate $\Delta{u}$ as:
\begin{equation}
\widehat{\Delta{u}}=\sqrt[4]{ \frac{G_{0,p}^{2} \gamma_{2}^{2} d_{ref}^{4}\delta{(t-(\frac{d}{c}))}^{2}}{\Delta{t}^{4}(a_1^{2}+b_1^{2})}}
\end{equation}

From the above equation, we can see that the estimated $\widehat{\Delta{u}}$ value depends on the excess delay time $\tau_{2}=\frac{d}{c}$ that is caused by the Doppler phenomenon.

\section{Performance Evaluation}

\label{performance_evaluation}
\subsection{Evaluation Setup}
\begin{figure}
\centering
 \includegraphics[keepaspectratio,width = 1\linewidth]{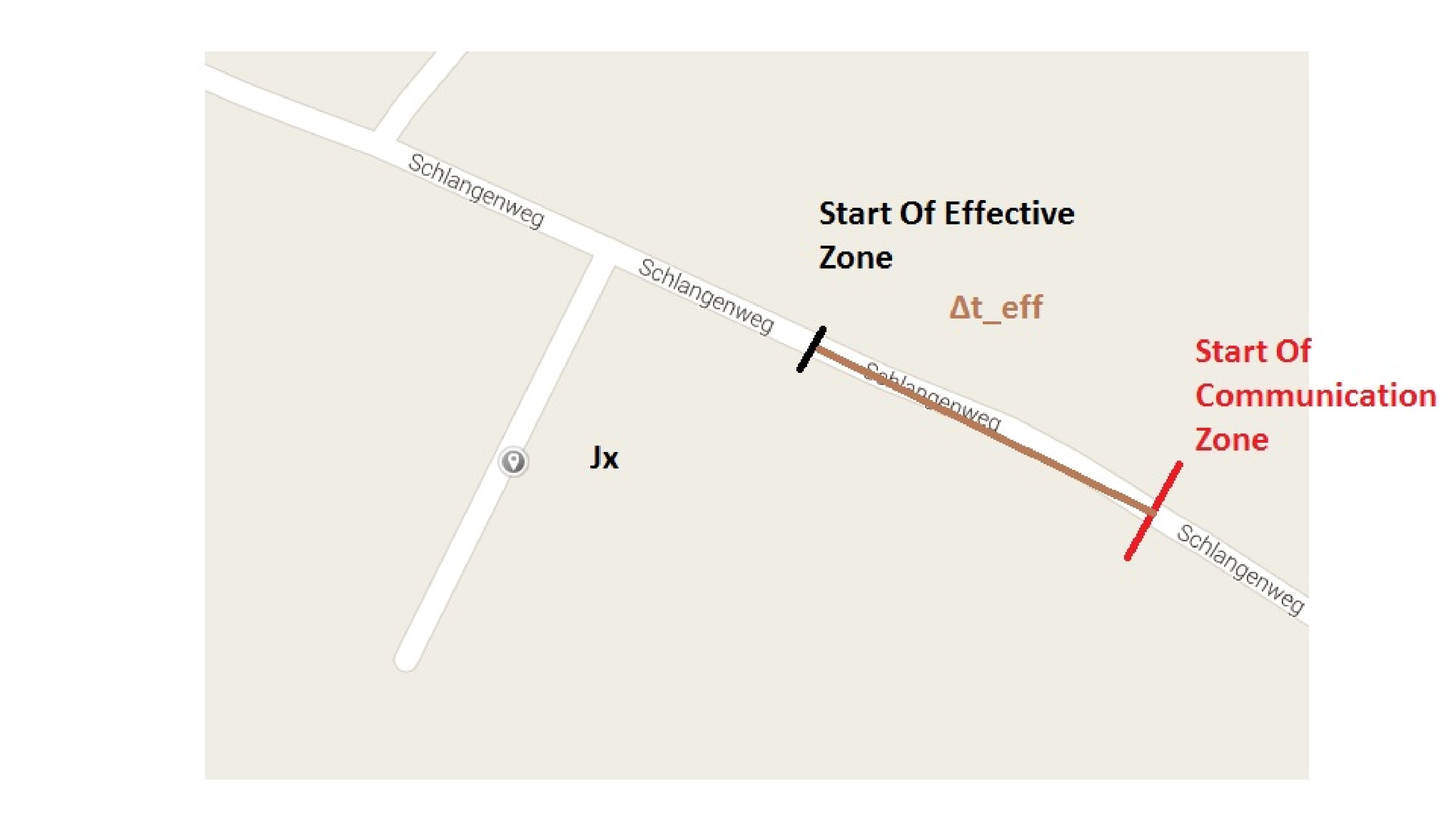}
\caption{Graphical Representation of the  $\Delta{t}_{eff}$ from the $Tx - Rx$ pair between the Communication Zone and Effective Zone of the $Jx$.}
\label{fig:situation-description}
\end{figure}

Our evaluation scenario is conducted on the outskirts of the city of Aachen, representing a real-word environment; while assuming that this is a rural area. Our experimental setup considers unicast data transmissions in a network consisting of three nodes: a transmitter, a receiver and a jammer, and V2X broadcast communication for 10 interfering vehicles outside of the CS range between the $Tx$ and the $Rx$ (distance more than 1000m).
Two different moving RF jamming attacks scenarios are evaluated. Analyzing RF Jammer Behavior 1 \ref{exp-1}, the $Tx$ - $Rx$ pair (see Fig.\ref{fig:doopler}) travels with a constant speed of approximately $50Km/h$ and with constant distance of approximately 20m, as a platoon of vehicles. The $J_x$ is also moving on a side road with zero initial speed and accelerates to a maximum speed of 60Km/h in order to approach the $Tx$ - $Rx$ pair. In RF Jammer Behavior 2 \ref{exp-2} the transmitter and the receiver travel with constant speed of approximately $48Km/h$ when the jammer approaches the crossroads, as illustrated in Fig.\ref{fig:situation-description}, with accelerating speed and a maximum limit of $50km/h$. 

Our experiments are conducted using the Veins-Sumo simulator \cite{veins} with the simulation parameters presented in Table~\ref{array:veins} such as: The initial distance between the jammer and the pair of $Rx$ - $Tx$, $d_{Jx-Rx}$, the distance that separates the receiver from the transmitter throughout the course of the simulation $d_{T_{x}-R_{x}}$. The closest distance in which the jammer arrives relative to the $Tx$ - $Rx$ pair as well as the power of all the transmitted signals $P_{T_{x},J_{x}}$. The time interval $\Delta{t}$ after RSEA is reapplied. The specific value of the parameter $N$, which is the number of the multipath rays. Last, the standard reference distance $d_{ref}$ is used for the estimation of the LOS path loss component.

As illustrated in Fig.\ref{fig:situation-description}, there is a time interval $\Delta{t}_{eff}$, in which the transmitter can effectively communicate with the receiver. It starts with the \textit{'Start of Communication Zone'} and ends with the \textit{'Start of the Effective Zone'} of the $Jx$. After the start of the effective zone of the $Jx$, the jammer is located at distances smaller than 30m away from the receiver and it can completely  jam the communication between the $Tx$ and the $Rx$ by constructing a \textit{'Black hole'}. All the evaluation parameters are summarized in Table~\ref{array:res}. 

During the performance evaluation we test our
proposed RSEA with different SINR values for two real-life scenarios. When the jamming vehicle is approaching the $Tx$ - $Rx$ pair, the SINR is:

\begin{equation}
SINR=\frac{||h_1\vec{x}_\text{pilot}||^{2}}{||h_2\vec{s}||^{2}+\sigma_{n}^{2}}
\end{equation}

The SINR level is measured by the receiver at the PHY layer. In the above equation, the noise power $\sigma_{n}^{2}$ is the noise power.
Moreover, the Mean Absolute Error (MAE) between the real $\Delta{u}$ value and the estimated is calculated for both scenarios. This is the difference between the actual relative speed metric $\Delta{u}$ with the estimated relative speed metric $\hat{\Delta{u}}$.
\begin{equation}
MAE=\frac{1}{ns}|\Delta{u}_{i}-\hat{\Delta{u}}_\text{i}|
\end{equation}
where $i$ is an integer number that identified with the current time instant in which the real and the estimated $\Delta{u}$ variable have a specific value and $ns=10$ is the number of measurements for the specific speed value. The MAE value gets its optimal zero value when the real $\Delta{u}$ is identified with the estimated. We assume this optimal value as a reference point for the MAE$(\%)$ calculations for the rest of the paper. 

\begin{center}
\begin{table}
\centering
\caption{Simulation Parameters}
\label{array:veins}
 \begin{tabular}{||c c||} 
 \hline
 \textbf{Evaluation Parameters in Veins Simulator}& \textbf{Values}  \\ [0.5 ex]
\hline\hline
 $d_{T_{x},R_{x}}$ & 20m  \\
 \hline
 $[CW[min], CW[max]]$ & [3,7] \\
 \hline
 Vehicle's Transmission Range & 130-300m \\
 \hline
 Initial $d_{Jx-Rx}$ & 300m \\
 \hline
 CS range of 802.11p protocol & 1000m\\
 \hline
 Interfering vehicles outside of CS range $Tx - Rx$ & 10\\
 \hline
 $d_{Jx-Rx}$ at "Black hole"  & 25m \\
 \hline
 $P_{Tx,Jx}$ & 100mW  \\ 
 \hline
 Minimum sensitivity $(P_{th})$ & -69dBm to -85dBm  \\
 \hline
 $f_c$ & 5.9GHz \\
  \hline
  Doppler shift for $\Delta{u}=120km/h$ & $\pm 655.5$ Hz  \\
  \hline
 $ d_{ref}$ & 100m \\
  \hline
  $\Delta{t}$ & 2s \\
  \hline
  $N$ & 4 \\
  \hline
 \end{tabular}
\end{table}
\end{center}

\subsection{Results of RF Jammer Behavior 1}
\label{exp-1}
\begin{figure*}
\centering
\begin{subfigure}{.4\textwidth}
 \includegraphics[keepaspectratio,width = 1.1\linewidth]{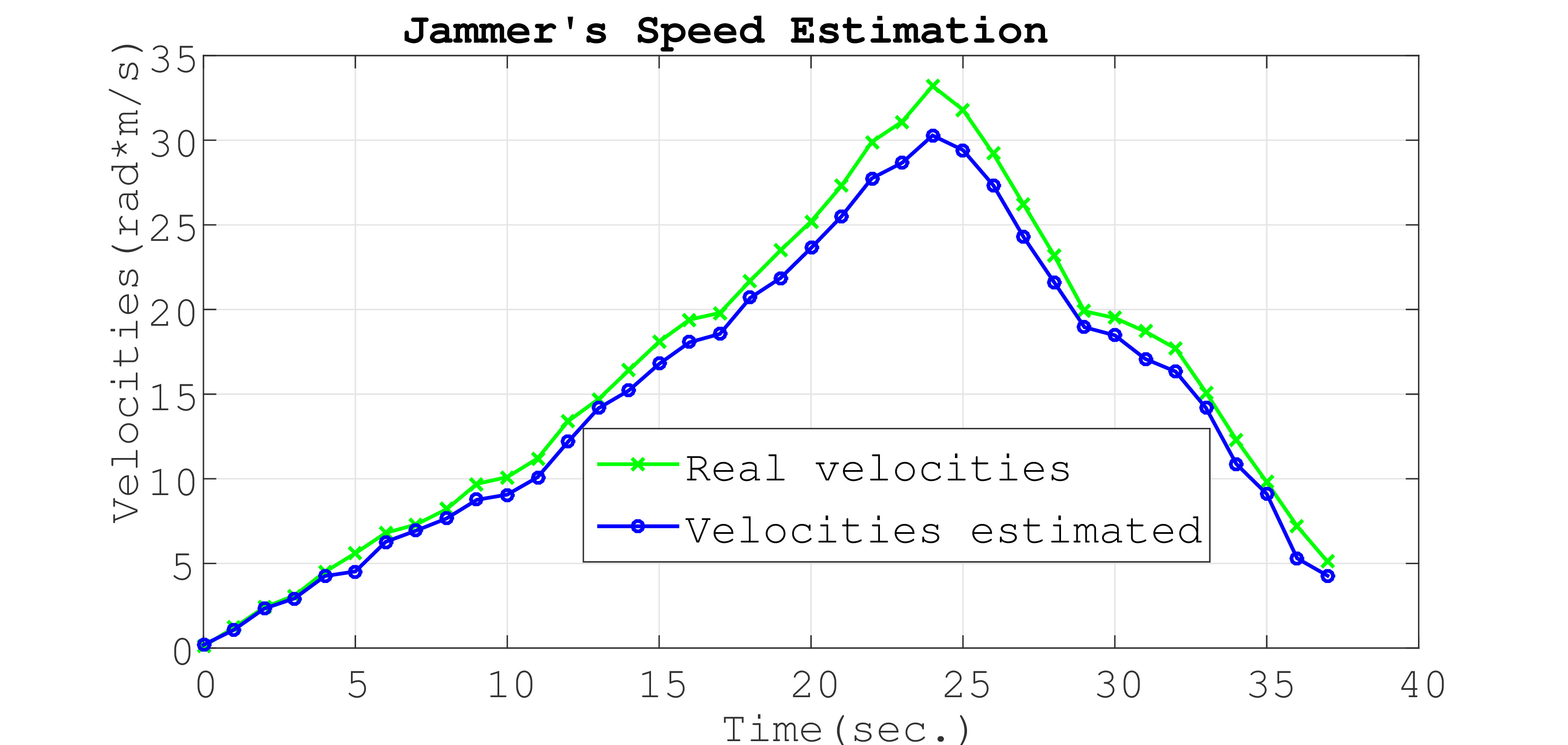}
 \caption{$\Delta{u}$ vs Estimated $\hat{\Delta{u}}$ to Time}
 \label{exp1:a}
 \end{subfigure}
 ~
 \begin{subfigure}{.5\textwidth}
 \includegraphics[keepaspectratio,width = 1\linewidth]{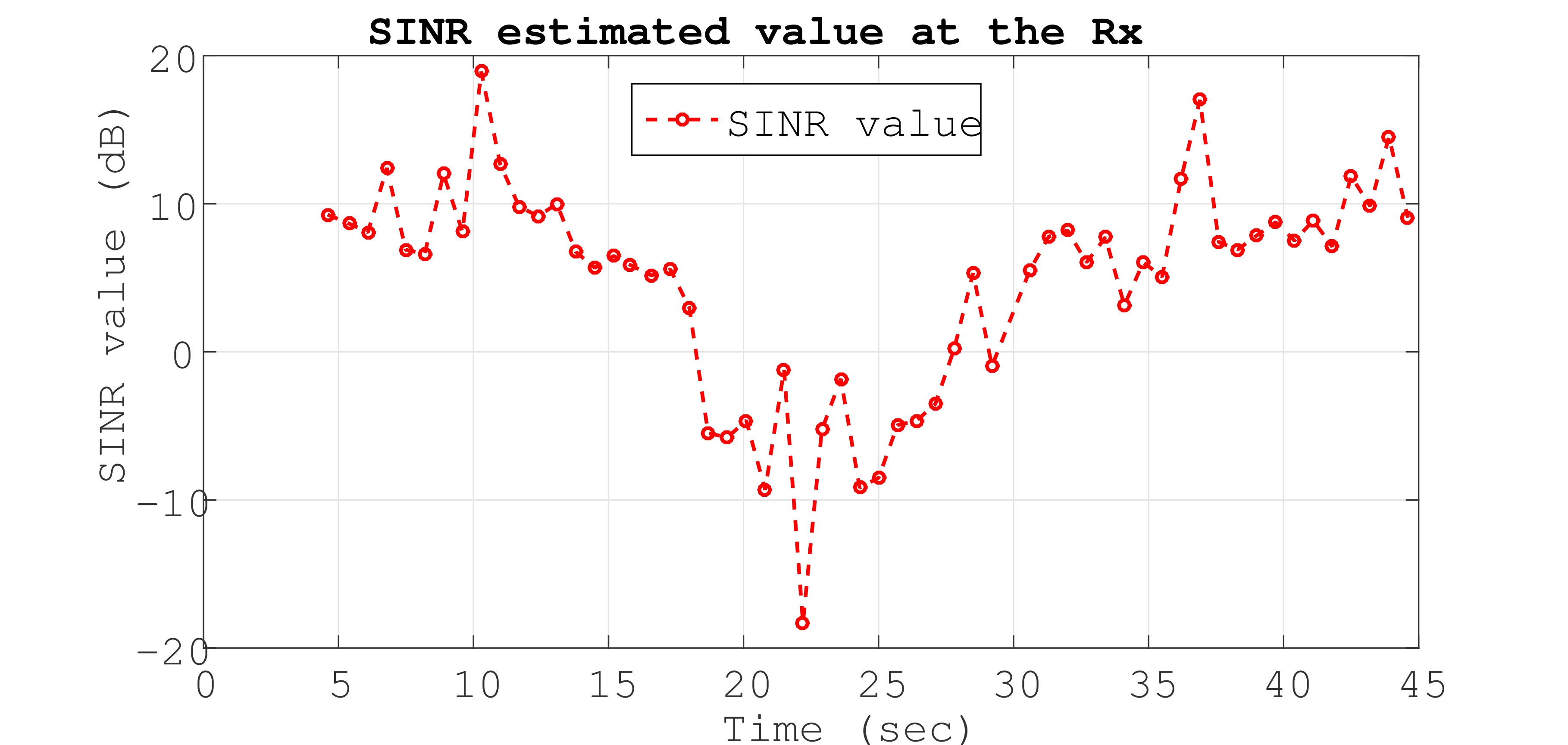}
  \caption{SINR(dB) vs Time(sec)}
   \label{exp1:b}
  \end{subfigure}

\caption{RF Jammer Behavior 1 results: RSEA with speed of $Tx$ - $Rx$ (50 Km/h) and $Jx$ maximum speed (60 Km/h) and a \textit{simplified} form of  jamming signal}
\label{fig:same_velocity}
\end{figure*}

In RF Jammer Behavior 1, we assume that the pair $Tx$ - $Rx$ moves with a high constant speed (50 Km/h) when the jammer accelerates with a higher maximum speed (60 Km/h), while transmitting
a jamming signal with a \textit{simplified} form to the receiver. The first figure of Fig.\ref{exp1:a} shows a comparison between the real $\Delta{u}$ and the estimated value. Specifically, by observing the start time of the steep slope of SINR in Fig.\ref{exp2:b}, we can conclude that it  coincides with the start of the jamming attack, the 15.5 sec. The main reason for the sharp decrease of the SINR in our experiment is the jamming attack and not the interference from the entire environment. Moreover, in Fig.\ref{exp1:a}, after 15.5 sec, for which $\Delta{u}$ is above $20$ rad*m/s, the SINR in Fig.\ref{exp1:b} has also a steep slope. So, the effective zone of communication between $Tx$ and $Rx$ is approximately $\Delta{t_{eff}}=15.5$ secs, whilst after that it is corrupted for 13.5 secs. So, the 'black-hole' in the communication range between the $Tx$ and the $Rx$ is during the time interval (15.5 secs- 29 secs). After 29 secs, we have the end of the attack. For this time interval (15.5 secs- 29 secs), the MAE of our proposed RSEA increases to $23\%$ from the optimal MAE value (see Fig \ref{fig:rse_comp}).

In Fig.\ref{exp1:a}, we can see that $\Delta{u}$ reaches a maximum value, approximately $32.5$ rad* m/s, at the time instant $23.4$ sec. At this time, $Jx$ is approaching the $Tx$ - $Rx$ pair in the main road, which is illustrated in Fig.\ref{fig:situation-description}. The average MAE for the duration of RF Jammer Behavior 1 is approximately $13\%$ worse than the optimal value. 

\begin{center}
\begin{table}

\centering
\caption{Evaluation scenario parameters}
\label{array:res}
 \begin{tabular}{||c c c||} 
 \hline
 Independent parameters & RF jammer 1 & RF jammer 2  \\ [0.5 ex]
\hline\hline
 $Tx$ - $Rx$ velocity & 50 Km/h & 48 Km/h \\ 
 \hline
 $Jx$ velocity & 60 Km/h & 50 Km/h \\
 \hline
 $\Delta{t}_{eff}$ & 15.5 sec & 18 sec  \\
\hline
 "Black hole" of communication & 13.5 sec & 18 sec \\
  \hline
  Time of $\Delta{u}$ peak & 23.4 sec & 25 sec \\
  \hline
 \end{tabular}
\end{table}
\end{center}

\subsection{Results of RF Jammer Behavior 2}
\label{exp-2}
\begin{figure*}
\centering
\begin{subfigure}{.4\textwidth}
 \includegraphics[keepaspectratio,width = 1.1\linewidth]{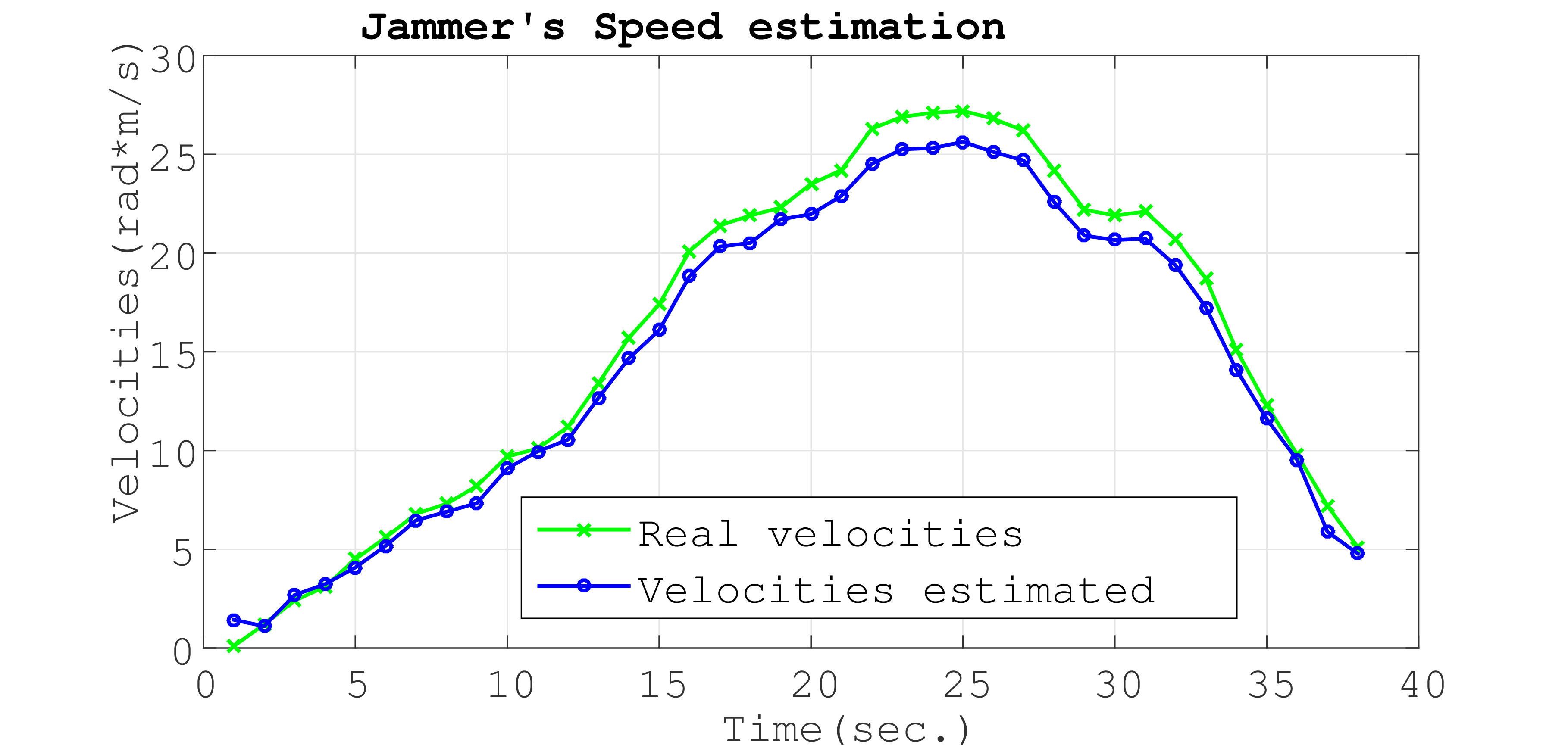}
  \caption{Real $\Delta{u}$ vs Estimated $\hat{\Delta{u}}$ to Time }
  \label{exp2:a}
 \end{subfigure}
 \begin{subfigure}{.5\textwidth}
  \includegraphics[keepaspectratio,width = 1\linewidth]{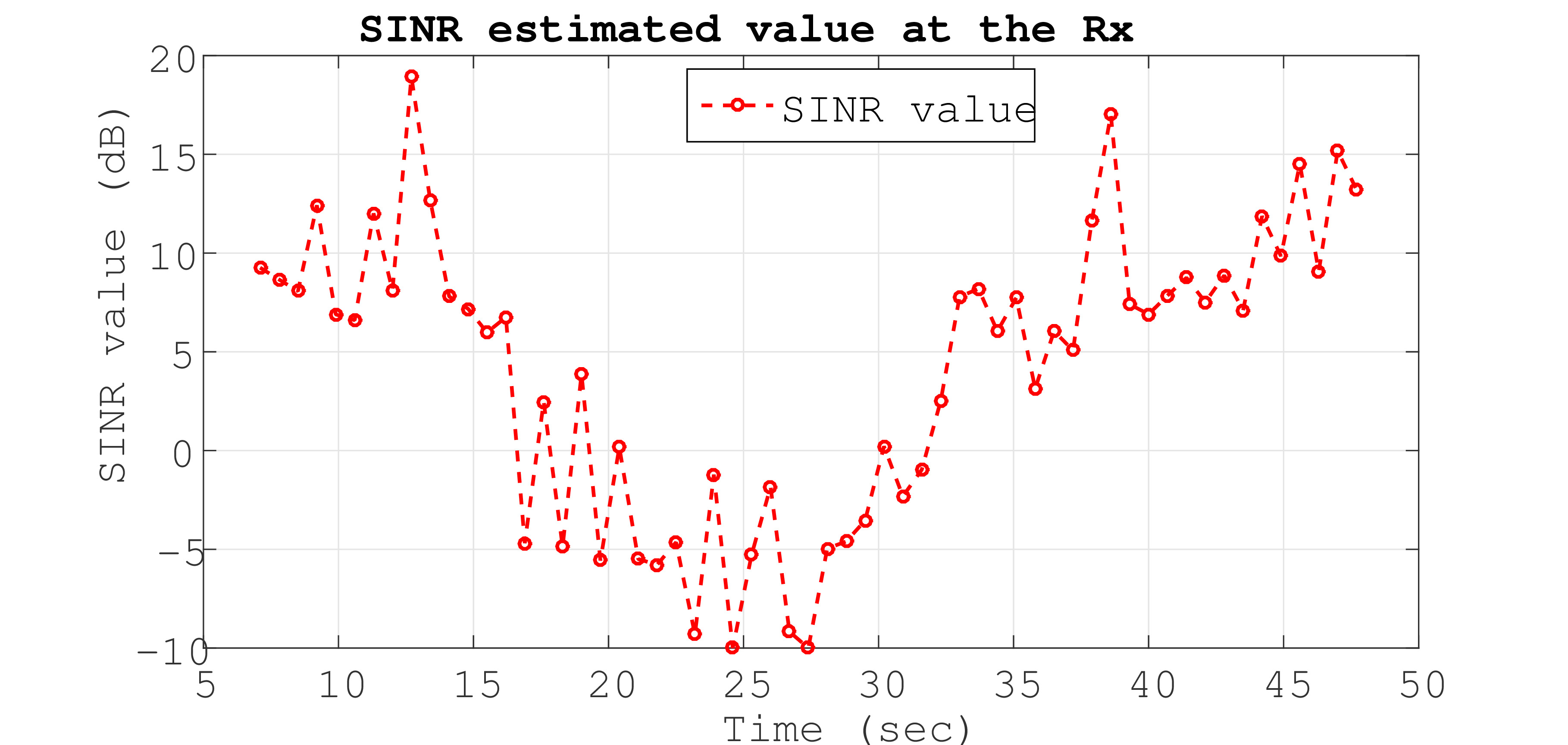}
  \caption{SINR(dB) vs Time(sec)}
  \label{exp2:b}
  \end{subfigure}
\caption{RF Jammer Behavior 2: RSEA with speed of $Tx$ - $Rx$ (48 Km/h) and $Jx$ maximum speed (50 Km/h) and \textit{simplified jamming signal} form}
\label{fig:diff_velocity}
\end{figure*}

For the second evaluation scenario, we assume that the pair $Tx$ - $Rx$ travels with constant speed (48 Km/h), which is almost the same as the maximum speed of the jammer (50 Km/h) (see Fig.\ref{fig:diff_velocity}). The jammer continuously transmits a random jamming symbol to the receiver. The start time of the jamming attack is at 18 secs during which the SINR appears to be decreasing from $5$dB to zero while $\Delta{u}$ starts to increase from $20$ rad*m/s to the 'peak' value of $\Delta{u}$. The time that is needed for the $Jx$ to approach the pair $Tx$ - $Rx$ is approximately $\Delta{t}_{eff}=18$ secs. After that time, the jamming attack clearly has perfect results for $18$ secs; from the $18$ secs of the simulation until $36$ secs, after that SINR increases more than $5$ dB. \par
If the $\Delta{u}$ slope is positive, the $Jx$ approaches the $Rx$, whilst if it goes to zero, $Jx$ is removed from the effective zone of communication between the $Tx$ and the $Rx$. 
The 'black-hole' in the communication between the $Tx$ and the $Rx$ is around the time interval (18 secs - 36 secs), during which the MAE value increases to approximately $18\%$ from the optimal MAE value.

In Fig.\ref{fig:diff_velocity}, we can see that the average MAE for the complete duration of RF Jammer Behavior 2 is approximately $10\%$ worse than the optimal value, as it is shown also in Fig.\ref{fig:rse_comp}

\subsection{MAE comparison between RF Jammer Behavior 1 and RF Jammer Behavior 2}

\begin{figure}
\centering
 \includegraphics[keepaspectratio,width = 1\linewidth]{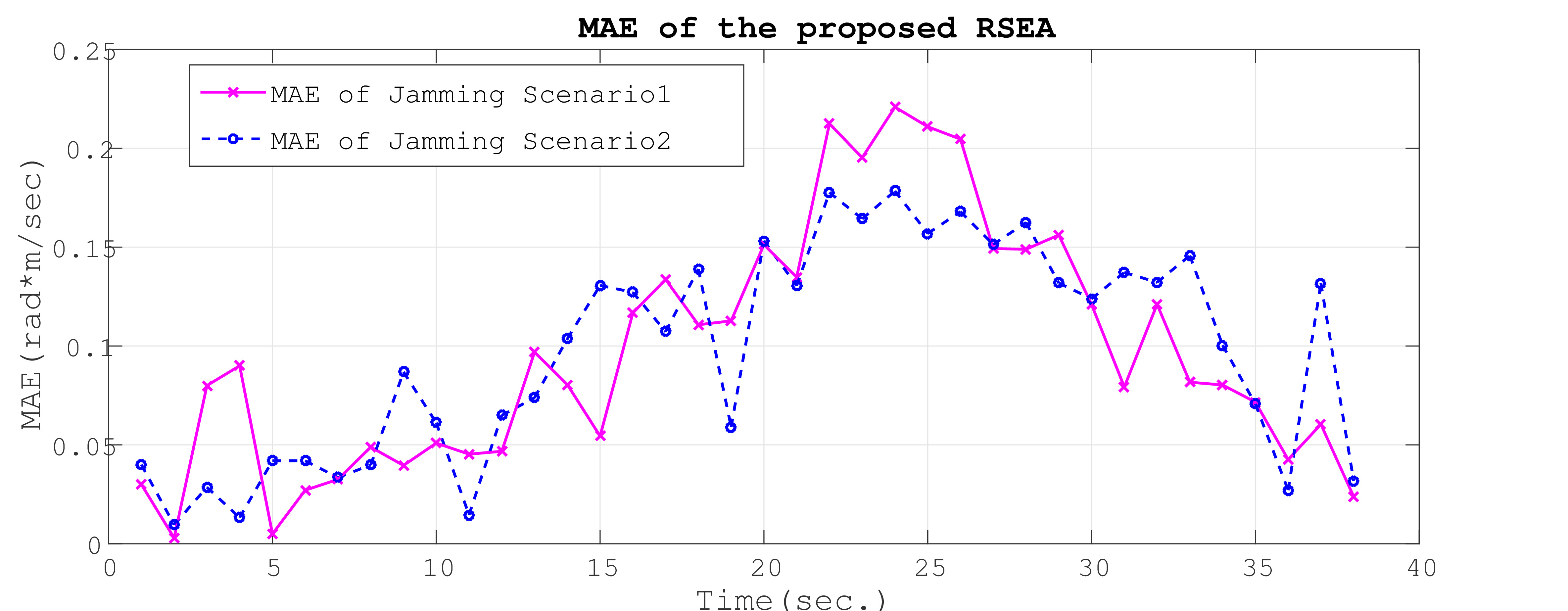}
  \caption{Instantaneous MAE comparison between RF Jammer Behavior 1 and RF Jammer Behavior 2 with \textit{simplified jamming signal} form}
\label{fig:rse_comp}
\end{figure}

The overall comparison of the MAE results between RF Jammer Behavior 1 (Jammer 1) and RF Jammer Behavior 2 (Jammer 2) is summarized in Table \ref{array:res_comp}. Fig.\ref{fig:rse_comp} shows that there is a quite small MAE, only $15\%$ greater than the MAE value at the start and end of simulation. However, when the jammer approaches the receiver the MAE shows an increase of about $23\%$ from the optimal value for RF Jammer Behavior 1 and $18\%$ for RF Jammer Behavior 2. The phenomenon of the larger MAE at the time of the jamming attack for RF Jammer Behavior 1 compared to that of RF Jammer Behavior 2 is attributed to the fast varying nature of the $\Delta{u}$ metric, because it changes with a higher rate and thus, the channel between the $Jx$ and the $Rx$ changes frequently too. So, the longer the duration of the jamming attack lasts the better the MAE results of the proposed RSEA are.   

For the duration of the effective "Communication Zone" between $Tx$ - $Rx$ $(\Delta{t}_{eff})$ the average MAE was only $8\%$ worse than the zero MAE value for RF Jammer Behavior 1 and $6\%$ for RF Jammer Behavior 2. As the jammer approaches the receiver at a distance of approximately 30m, it could successfully affect the effective communication between $Tx - Rx$. For the time interval of a "Black hole" in communication between $Tx$ - $Rx$, the average MAE increases to $23\%$ from the optimal value for RF Jammer Behavior 1 and $18\%$ for RF Jammer Behavior 2. When the overall average MAE, for the duration of the simulation is measured, an average MAE $13\%$ worse than the zero MAE value for RF Jammer Behavior 1 is observed and $10\%$ worse than the same reference point for RF Jammer Behavior 2. An average MAE for our scheme for all scenarios can be estimated and is about $13\%$ worse than the optimal MAE value. Finally, it can be concluded from the simulation results that neither the jamming signal form nor the different $Jx$ and $Tx - Rx$ pair speeds affect the speed estimation results significantly. Moreover, the insertion of 10 "hidden" nodes further away from the CS range of  $Tx - Rx$ does not seem to affect the speed estimation values that are presented. A "hidden" node occupies the wireless medium through the MAC/EDCA backoff mechanism (see \ref{mac-pilot}), in very rare cases. In these cases, the $Tx - Rx$ communication is disrupted, forcing the RSEA to use the signal that $Rx$ receives from this node for the combined channel estimation and finally, for the $\Delta{u}$ estimation.

\begin{figure*}
\centering
\begin{subfigure}{.4\textwidth}
 \includegraphics[keepaspectratio,width = 1.1\linewidth]{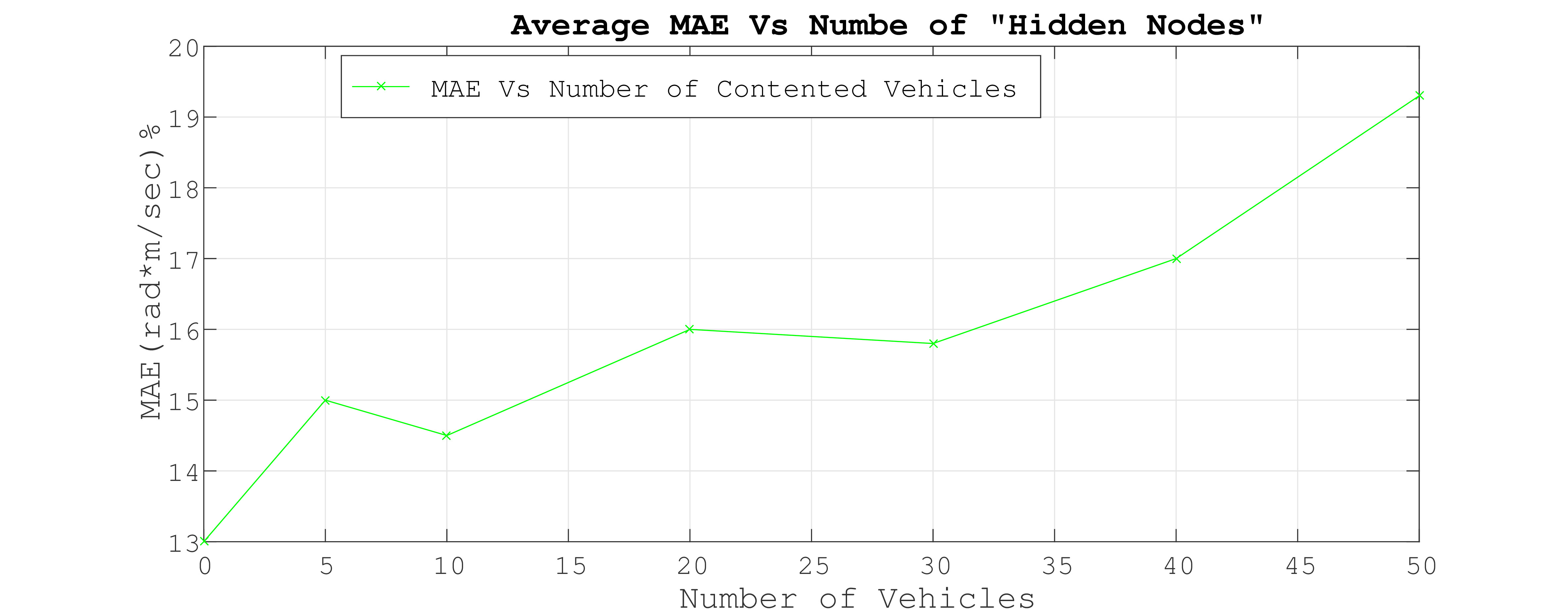}
  \caption{Average MAE $(\%)$ vs Different number of "hidden" nodes}
  \label{exp3:a}
 \end{subfigure}
 \begin{subfigure}{.5\textwidth}
  \includegraphics[keepaspectratio,width = 0.9\linewidth]{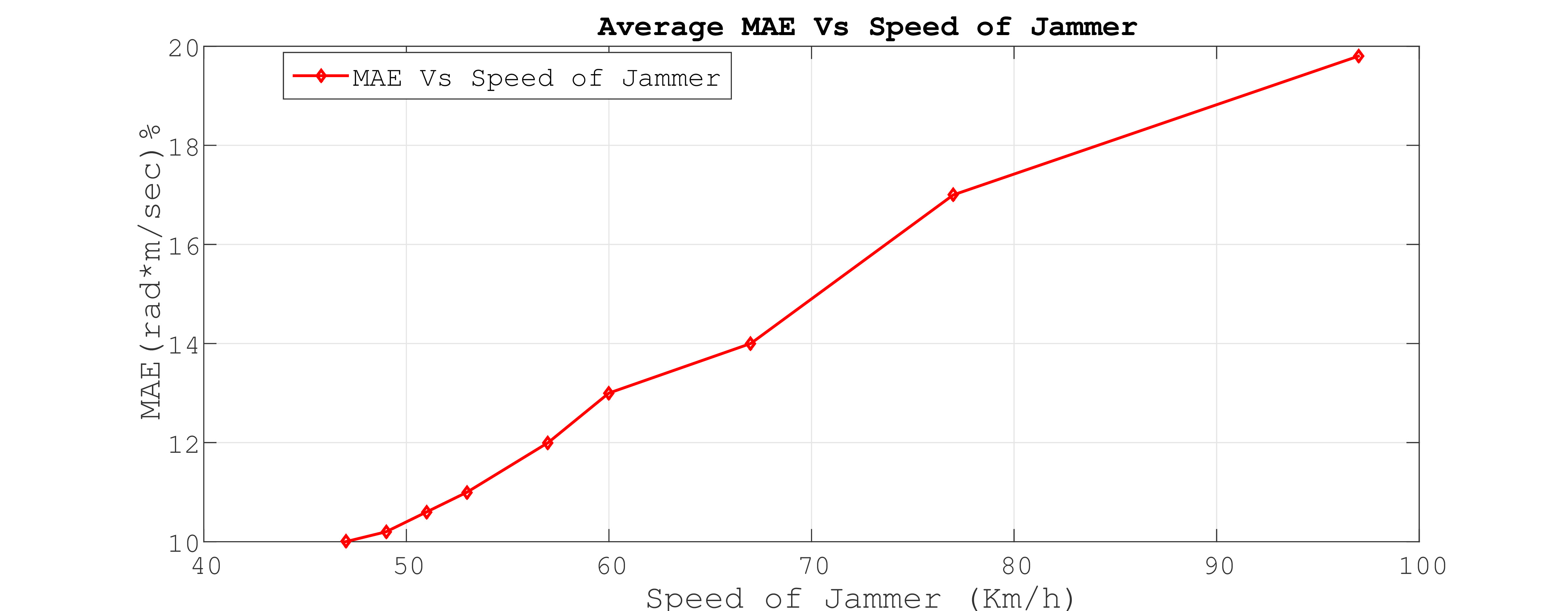}
  \caption{Average MAE $(\%)$ vs Range of jammer's speed values(Km/h)}
  \label{exp3:b}
  \end{subfigure}

\caption{Average MAE$(\%)$ increase from the optimal zero MAE reference point with different evaluation parameters: The MAE of the proposed RSEA with different number of nodes for the contention window of the MAC backoff procedure for the wireless channel bettween $Tx$ - $Rx$ and different jammer's speed values.}
\label{fig:diff_velocity_backoff}
\end{figure*}

In order to test our previous results under more generic scenarios, the average MAE is estimated for different jammer speed values and different number of "hidden" nodes that are located at the edge of the CS range of the $Tx - Rx$ pair. Specifically we conducted several simulations, for a range of jammer speed values between $[47,97]Km/h$ and number of "hidden" nodes between $[0,50]$ nodes. For these parameters, the MAE value increases at approximately $20\%$ from the reference zero MAE value with the maximum jammer speed value (see Fig.\ref{exp3:b}) and at approximately $19,2\%$ from the same reference value with the maximum number of "hidden" nodes, which is 50 nodes (see Fig.\ref{exp3:a}). For values greater than $67km/h$ regarding the speed metric and $30$ "hidden" nodes, the MAE was increasing with a higher rate. For smaller values of these two "side- effect" values, the increase of MAE value is negligible. So these simulation results indicate that the backoff MAC/EDCA algorithm, using a safety-related high priority channel for the communication between $Tx - Rx$, does not affect considerably the performance of the speed estimation algorithm. The jammer's speed increase, also, affects but not significantly the proposed RSEA. 

\begin{center}
\begin{table}

\centering
\caption{RF Jammer Behavior comparison results of MAE$(\%)$ increase from the optimal zero MAE reference point.}
\label{array:res_comp}
 \begin{tabular}{||c c c||} 
  \hline
Time Intervals & MAE (Jammer 1) &MAE (Jammer 2)  \\ [0.5 ex]
\hline\hline
 "Black hole" of communication $Tx$ - $Rx$ & $23\%$ & $18\%$ \\
  \hline
  Time Interval of $\Delta{t}_{eff}$ & $8\%$ & $6\%$ \\
  \hline
  Overall Simulation Time   & $13\%$ & $10\%$ \\
  \hline
 \end{tabular}
\end{table}
\end{center}

\section{Discussion}

In Section \ref{performance_evaluation} we tested our proposed RSEA under different SINR values in order to represent realistic conditions. V2X communication generally uses broadcast messages, but in this paper we use unicast RF communication between two nodes  in order to perform jammer's speed estimation. This type of communication is supported for advanced safety applications of autonomous vehicles by the Qualcomm's Cellular Vehicle-to-Everything (C-V2X) technology \cite{qualcom}. The target of this paper is to evaluate the performance of the RSEA for a pair of moving nodes with limited conditions, having as a future objective to be used in a real-life VANET scenario for more than one pair of nodes. Peer to peer networks in vanets are studied lately in many other works \cite{ferrag2017esspr,kumar2014peer, rezende2015reactive}, focusing mostly on social networks message exchange, cooperative caching or unicast video streaming.

Relative speed estimation results from our proposed RSEA can be collected from a Trusted Central Authority (TCA) that exists in the area. Analyzing these collected data records, the TCA make deductions based on the SINR value, notify approaching vehicles and even propose jamming free routes \cite{trusted-routing}. 

\section{Conclusion and Future Work}
\label{conclusion}
In this paper, we presented an algorithm for estimating the combined value $\Delta{u}$ of the relative speed between $Rx$ and $Jx$ in combination with the AOP of the $Jx$, during a jamming attack. A \textit{simplified jamming signal} is sent to the receiver by the jammer, that contains the same unknown symbol to the receiver $K$ times. An entirely unknown jamming signal can also be estimated by the receiver at the proposed algorithm, too. The proposed relative speed metric can capture both the speed of the jammer and its direction relative to the $Tx$ - $Rx$ movement. By predicting the above value, we can understand jammer's behavior, for which $Rx$ does not have any information except for the combined signal that is received from $Tx$ and the interference caused by the attacker. Our proposed RSEA uses the physical metric of $\Delta{u}$ from RF communication $Tx - Rx$ in order to estimate the direction of the attacker. This metric is combined with the SINR value from the hardware (physical layer) in order for a real-life VANET scenario to be simulated. The MAE measured is being approximately only $10\%$  worse compared to the optimal zero MAE value under different jamming attach scenarios.

As future work, we plan to combine RSEA with other metrics from the PHY layer or the network layer, such as SINR, for developing an accurate cross-layer jamming detection scheme. The detection scheme will be capable to deal with more than one pair of nodes that communicate in a broadcast form. This combined metric can be also used as an extra feature in a machine-learning approach (see \cite{machine-learning}), in which the vehicles of the area can be classified as cooperative or malicious, thereby forming a trusted vehicular network. The usage of the relative speed metric can also reduce false alarms and can provide additional information about the future position of a $Jx$, such as the time that the attacker will approach the effective zone of communication. The above information extracted from our channel based $Jx$ - $Rx$ analysis, can decrease false alarms compared to jamming prediction schemes that are based only on the 802.11p PHY/MAC related metrics (see the DJAVAN in \cite{djavan}), concluding the physical geographical topology of the attacker. Last but not least, this metric is appropriate for a variety of jamming attacks.

\bibliographystyle{IEEEtran}
\bibliography{Template} 

\begin{thebibliography}{10}
\providecommand{\url}[1]{#1}
\csname url@samestyle\endcsname
\providecommand{\newblock}{\relax}
\providecommand{\bibinfo}[2]{#2}
\providecommand{\BIBentrySTDinterwordspacing}{\spaceskip=0pt\relax}
\providecommand{\BIBentryALTinterwordstretchfactor}{4}
\providecommand{\BIBentryALTinterwordspacing}{\spaceskip=\fontdimen2\font plus
\BIBentryALTinterwordstretchfactor\fontdimen3\font minus
  \fontdimen4\font\relax}
\providecommand{\BIBforeignlanguage}[2]{{%
\expandafter\ifx\csname l@#1\endcsname\relax
\typeout{** WARNING: IEEEtran.bst: No hyphenation pattern has been}%
\typeout{** loaded for the language `#1'. Using the pattern for}%
\typeout{** the default language instead.}%
\else
\language=\csname l@#1\endcsname
\fi
#2}}
\providecommand{\BIBdecl}{\relax}
\BIBdecl

\bibitem{platoons}
A.~Mani, R.~Arun, C.~Chen-Nee, and G.~Dipak, ``Security vulnerabilities of
  connected vehicle streams and their impact on cooperative driving,''
  \emph{IEEE Communications Magazine}, vol.~53, pp. 126--132, 2015.

\bibitem{guardian}
A.~Herm, ``{Assume self-driving cars are a hacker's dream? Think again },''
  \url{https://www.theguardian.com/technology/2017/aug/30/self-driving-cars-hackers-security},
  2017, [Online; accessed 30-August-2017].

\bibitem{new-york-times}
D.~WAKABAYASHI, ``{Waymo's Autonomous Cars Cut Out Human Drivers in Road
  Tests},''
  \url{https://www.nytimes.com/2017/11/07/technology/waymo-autonomous-cars.html},
  2017, [Online; accessed 7-November-2017].

\bibitem{vehicle-platooning}
C.~Darren, ``{What is vehicle platooning? Driving Tests},''
  \url{https://www.drivingtests.co.nz/resources/what-is-vehicle-platooning/},
  2017.

\bibitem{ivg}
J.~A. Hartigan and M.~A. Wong, ``Internet of vehicles: From intelligent grid to
  autonomous cars and vehicular clouds,'' \emph{Internet of Things (WF-IoT),
  2014 IEEE World Forum}, 2014.

\bibitem{iov2017}
J.~Jo and M.~Gerla, ``Internet of vehicles and autonomous connected car -
  privacy and security issues,'' \emph{Computer Communication and Networks
  (ICCCN), 2017 26th International Conference}, 2017.

\bibitem{wifi-car}
R.~Barrett, ``{Wi-Fi} in the car: how to meet the concurrent needs of multiple
  systems and applications,''
  \url{http://www.eenewseurope.com/design-center/wi-fi-car-how-meet-concurrent-needs-multiple-systems-and-applications-0},
  2017.

\bibitem{quyoom2015novel}
A.~Quyoom, R.~Ali, D.~N. Gouttam, and H.~Sharma, ``A novel mechanism of
  detection of denial of service attack ({DoS}) in {VANET} using malicious and
  irrelevant packet detection algorithm ({MIPDA}),'' \emph{Computing,
  Communication \& Automation (ICCCA), 2015 International Conference on}, pp.
  414--419, 2015.

\bibitem{YOUSAF2017124}
A.~Yousaf, A.~Loan, R.~F. Babiceanu, L.~Maglaras, and O.~Yousaf,
  ``Architectural and information theoretic perspectives of physical layer
  intruders for direct sequence spread spectrum systems,'' \emph{Computers and
  Security}, 2017.

\bibitem{velocity2009cell}
Z.~Ehsan and A.~Ghasem, ``{IF}-based velocity estimation of the mobile units in
  micro-cellular systems with non-isotropic scattering distribution,''
  \emph{Vehicular Technology Conference Fall (VTC 2009-Fall), 2009 IEEE 70th},
  2009.

\bibitem{jamming-survey}
S.~Malebary and W.~Xu, ``A survey on jamming in vanet,'' \emph{International
  Journal of Scientific Research and Innovative Technology}, vol.~2, no.~1,
  2015.

\bibitem{rf-jamming-punal}
O.~Punal, C.~Pereira, A.~Aguiar, and J.~Gross, ``Experimental characterization
  and modeling of {RF} jamming attacks on {VANETs},'' \emph{IEEE Transactions
  on Vehicular Technology}, vol.~64, pp. 524 -- 540, 2015.

\bibitem{reactive2}
M.~Pajic and R.~Mangharam, ``Spatio-temporal techniques for antijamming in
  embedded wireless networks,'' \emph{EURASIP Journal on Wireless
  Communications and Networking}, vol. 2010, 2010.

\bibitem{reactive3}
A.~Wood, J.~Stankovic, and G.~Zhou, ``{DEEJAM}: Defeating energyefficient
  jamming in {IEEE} 802.15.4-based wireless networks,'' \emph{Sensor, Mesh and
  Ad Hoc Communications and Networks, 2007. SECON '07. 4th Annual IEEE
  Communications Society Conference on}, June 2007.

\bibitem{reactive1}
M.~Wilhelm, I.~Martinovic, J.~B. Schmitt, and V.~Lenders, ``Short paper:
  reactive jamming in wireless networks: how realistic is the threat?''
  \emph{Proceedings of the fourth ACM conference on Wireless network security},
  2011.

\bibitem{reactive4}
D.~Nguyen, C.~Sahin, B.~Shishkin, N.~Kandasamy, and K.~R. Dandekar, ``A
  real-time and protocol-aware reactive jamming framework built on
  software-defined radios,'' \emph{Proceedings of the 2014 ACM Workshop on
  Software Radio Implementation Forum, ser. SRIF '14. New York, NY, USA:
  ACM}, 2014.

\bibitem{reactive5}
E.~Bayraktaroglu, C.~King, X.~Liu, G.~Noubir, R.~Rajaraman, and B.~Thapa, ``On
  the performance of ieee 802.11 under jamming,'' \emph{NFOCOM, 2008
  Proceedings IEEE}, April 2008.

\bibitem{reactive6}
A.~Marttinen, A.~Wyglinski, and R.~Jantti, ``Statistics-based jamming detection
  algorithm for jamming attacks against tactical {MANETs},'' \emph{Military
  Communications Conference (MILCOM), 2014 IEEE}, Oct 2014.

\bibitem{reactive7}
M.~Strasser, B.~Danev, and S.~Capkun, ``Detection of reactive jamming in sensor
  networks,'' \emph{ACM Trans. Sen. Netw}, vol.~7, no.~2, 2010.

\bibitem{ofdm-pilot-jamming}
T.~C. Clancy, ``Efficient ofdm denial: Pilot jamming and pilot nulling,''
  \emph{2011 IEEE International Conference on Communications (ICC)}, 2011.

\bibitem{power-allocation}
S.~D'Oro, E.~Ekici, and S.~Palazzo, ``Optimal power allocation and scheduling
  under jamming attacks,'' \emph{IEEE/ACM Transactions on Networking}, vol.~25,
  pp. 1310 -- 1323, 2017.

\bibitem{uhf-infocom}
K.~Xu, Q.~Wang, and K.~Ren, ``Joint ufh and power control for effective
  wireless anti-jamming communication,'' \emph{2012 Proceedings IEEE INFOCOM},
  2012.

\bibitem{uhf-friendly-jammers}
J.~S. Sousa and J.~P. Vilela, ``Uncoordinated frequency hopping for secrecy
  with broadband jammers and eavesdroppers,'' \emph{2015 IEEE International
  Conference on Communications (ICC)}, 2015.

\bibitem{velocity-estimation}
T.~R. Raj, L.~Geert, and v.~d.~V. Alle-Jan, ``Relative velocity estimation
  using multidimensional scaling,'' \emph{Computational Advances in
  Multi-Sensor Adaptive Processing (CAMSAP), 2013 IEEE 5th International
  Workshop}, 2013.

\bibitem{speed-etc}
W.~Jiancheng, Y.~Rongliang, W.~Ru, and W.~Chang, ``Freeway travel speed
  calculation model based on {ETC} transaction data,'' \emph{Computational
  Intelligence and Neuroscience}, vol. 2014, pp. 514 -- 517, 2014.

\bibitem{speed-doopler}
K.~Branislav, L.~Akos, and K.~Xenofon, ``Tracking mobile nodes using {RF}
  {Doppler} shifts,'' \emph{SenSys '07 Proceedings of the 5th international
  conference on Embedded networked sensor systems}, pp. 29--42, 2007.

\bibitem{speed-estimation}
E.~A. Ahmad, K.~Markus, and F.~Georg, ``Distance and vehicle speed estimation
  in {OFDM} multipath channels,'' \emph{Microwave, Radar and Wireless
  Communications (MIKON), 2016 21st International Conference}, 2016.

\bibitem{reVISE}
K.~Nehal, K.~Ahmed~E., and Y.~Moustafa, ``{RF}-based vehicle detection and
  speed estimation,'' \emph{Vehicular Technology Conference (VTC Spring), 2012
  IEEE 75th}, 2012.

\bibitem{MUSIC}
B.~Wu, ``Realization and simulation of {DOA} estimation using {MUSIC} algorithm
  with uniform circular arrays,'' \emph{The 2006 4th Asia-Pacific Conference
  (2006)}, pp. 908--912, 2016.

\bibitem{speed-spectral}
C.~Tepedelenlioglu and G.~Giannakis, ``A spectral moment approach to velocity
  estimation in mobile communications,'' \emph{Wireless Communications and
  Networking Confernce (WCNC), IEEE, 2000}, 2000.

\bibitem{mobile-speed}
R.~Z. Yahong and X.~Chengshan, ``Mobile speed estimation for broadband wireless
  communications over rician fading channels,'' \emph{IEEE Transactions on
  Wireless Communications}, vol.~8, pp. 1--8, 2009.

\bibitem{speed-mobile-phones}
C.~Gayathri, V.~A.~V. Tam, G.~Marco, P.~M. Richard, Y.~Jie, and C.~Yingying,
  ``Vehicular speed estimation using received signal strength from mobile
  phones,'' \emph{UbiComp '10 Proceedings of the 12th ACM international
  conference on Ubiquitous computing}, pp. 237--240, 2010.

\bibitem{speed-mobile-phones-new}
P.~K. Pedapolu, P.~Kumar, V.~Harish, S.~Venturi, S.~K. Bharti, V.~Kumar, and
  S.~Kumar, ``Mobile phone user's speed estimation using {WiFi}
  {Signal-to-Noise Ratio},'' in \emph{Proceedings of the 18th ACM International
  Symposium on Mobile Ad Hoc Networking and Computing}.\hskip 1em plus 0.5em
  minus 0.4em\relax ACM, 2017, p.~32.

\bibitem{tracking}
K.~Branislav, L.~Akos, and K.~Xenofon, ``{SNR}-independent velocity estimation
  for mobile cellular communications systems,'' \emph{Proceeding SenSys '07
  Proceedings of the 5th international conference on Embedded networked sensor
  systems}, pp. 29--42, 2007.

\bibitem{angle-of-arrival}
A.~Abdelaziz, C.~E. Koksal, and H.~El~Gamal, ``On the security of angle of
  arrival estimation,'' in \emph{Communications and Network Security (CNS),
  2016 IEEE Conference on}, 2016, pp. 109--117.

\bibitem{doopler-shift}
A.~J. Javad and A.~G. Seyed, ``Doppler shift estimation and jamming detection
  for cellular networks,'' \emph{Australian Journal of Basic and Applied
  Sciences}, pp. 6590--6597, 2010.

\bibitem{detection-reactive}
S.~Michael, L.~Vincent, and M.~Wilhelm, ``Detection of reactive jamming in
  {DSSS}-based wireless communications,'' \emph{IEEE Transactions on Wireless
  Communications}, vol.~13, pp. 165 -- 171, 2014.

\bibitem{ddos-attack}
M.~Shabbir, M.~A. Khan, U.~S. Khan, and N.~A. Saqib, ``Detection and prevention
  of distributed denial of service attacks in {VANETs},'' in
  \emph{Computational Science and Computational Intelligence (CSCI), 2016
  International Conference on}.\hskip 1em plus 0.5em minus 0.4em\relax IEEE,
  2016, pp. 970--974.

\bibitem{karagiannis}
K.~Dimitrios and A.~Antonios, ``Jamming attack detection in a pair of rf
  communicating vehicles using unsupervised machine learning,'' in
  \emph{Vehicular Communications}, vol.~13.\hskip 1em plus 0.5em minus
  0.4em\relax ELSEVIER, 2018, pp. 56--63.

\bibitem{tse}
D.~Tse and P.~Viswanath, \emph{Fundamentals of wireless communication}.\hskip
  1em plus 0.5em minus 0.4em\relax Cambridge university press, 2005.

\bibitem{multipath-fading}
E.~Fatima, Abdeen~Farah, G.~E. Ashraf, and M.~H.~M. Nerma, ``A study of channel
  estimation in fast fading environments,'' in \emph{International Journal of
  Scientific and Technology Research}, vol.~4.\hskip 1em plus 0.5em minus
  0.4em\relax Directory of Open Access Journals, 2015, pp. 196--203.

\bibitem{GEMV-BOban}
B.~Mate, B.~João, and K.~T. Ozan, ``Geometry-based {Vehicle-to-Vehicle}
  channel modeling for large-scale simulation,'' \emph{IEEE Transactions on
  Vehicular Technology}, vol.~63, pp. 4146 -- 4164, 2016.

\bibitem{mmse}
S.~Aymen, C.~Faiza, K.~Lotfi, E.~Yassin, and R.~Atika, ``A symbol-based
  estimation technique for inter-vehicular communication performance
  optimization,'' \emph{IJCSI International Journal of Computer Science
  Issues}, vol.~10, 2014.

\bibitem{adaptive-model}
A.~Ghassan~M.T., A.-R. Mosa, and S.~Sidi-Mohammed, ``An adaptive channel model
  for {VBLAST} in vehicular networks,'' \emph{EURASIP Journal on Wireless
  Communications and Networking}, vol. 2009, 2009.

\bibitem{path_loss}
N.~Alam, A.~T. Balaie, and A.~G. Dempster, ``Dynamic path loss exponent and
  distance estimation in a vehicular network using doppler effect and received
  signal strength,'' \emph{Vehicular Technology Conference Fall (VTC
  2010-Fall), 2010 IEEE 72nd}, pp. 1--5, 2010.

\bibitem{stochastic1}
J.~Nuckelt, M.~Schack, and T.~Kurner, ``Deterministic and stochastic channel
  models implemented in a physical layer simulator for {Car-to-X}
  communications,'' \emph{Advances in Radio Science}, vol.~9", pp. 165 -- 171,
  2011.

\bibitem{veins}
S.~Christoph, G.~Reinhard, and F.~Dressler, ``Bidirectionally coupled network
  and road traffic simulation for improved {IVC} analysis,'' \emph{IEEE
  Transactions on Mobile Computing}, vol.~10, pp. 3--15, 2011.

\bibitem{qualcom}
QUALCOMM, ``Accelerating {C-V2X} commercialization,''
  \url{https://www.qualcomm.com/documents/path-5g-cellular-vehicle-everything-c-v2x},
  2017.

\bibitem{ferrag2017esspr}
M.~A. Ferrag and A.~Ahmim, ``{ESSPR}: an efficient secure routing scheme based
  on searchable encryption with vehicle proxy re-encryption for vehicular
  peer-to-peer social network,'' \emph{Telecommunication Systems}, vol.~66,
  no.~3, pp. 481--503, 2017.

\bibitem{kumar2014peer}
N.~Kumar and J.-H. Lee, ``Peer-to-peer cooperative caching for data
  dissemination in urban vehicular communications,'' \emph{IEEE Systems
  Journal}, vol.~8, no.~4, pp. 1136--1144, 2014.

\bibitem{rezende2015reactive}
C.~Rezende, A.~Boukerche, H.~S. Ramos, and A.~A. Loureiro, ``A reactive and
  scalable unicast solution for video streaming over {VANETs},'' \emph{IEEE
  Transactions on Computers}, vol.~64, no.~3, pp. 614--626, 2015.

\bibitem{trusted-routing}
N.~J. Patel and R.~H. Jhaveri, ``Trust based approaches for secure routing in
  {VANET}: A survey,'' \emph{Procedia Computer Science}, vol.~45, pp. 592--601,
  2015.

\bibitem{machine-learning}
P.~Oscar, A.~Ismet, S.~Caj-Julian, A.~Gloria, Ë.~W. Klaus, and G.~James,
  ``Machine learning-based jamming detection for {IEEE} 802.11: Design and
  experimental evaluation,'' \emph{World of Wireless, Mobile and Multimedia
  Networks (WoWMoM), 2014 IEEE 15th International Symposium}, 2014.

\bibitem{djavan}
M.~Lynda, B.-O. Jalel, and T.~N. Anh, ``{DJAVAN}: Detecting jamming attacks in
  vehicle ad hoc networks,'' \emph{Recent Advances in Modeling and Performance
  Evaluation in Wireless and Systems}, vol.~87, pp. 47--59, 2015.

\end{thebibliography}

\end{document}